\documentclass[referee, sn-nature]{sn-jnl}
\usepackage{geometry}
\usepackage{graphicx}%
\usepackage{amsmath,amssymb,amsfonts}%
\usepackage{amsthm}%
\usepackage{mathrsfs}%
\usepackage[title]{appendix}%
\usepackage{textcomp}%
\usepackage{manyfoot}%
\usepackage{booktabs}%
\usepackage{algorithm}%
\usepackage{algorithmicx}%
\usepackage{algpseudocode}%
\usepackage{listings}%
\usepackage{gensymb}
\usepackage{siunitx}
\usepackage{longtable}
\usepackage{seqsplit}
\usepackage{soul}
\usepackage{xcolor}
\usepackage[version=4]{mhchem}
\DeclareSIUnit{\base}{base}
\DeclareSIUnit{\molar}{M}
\DeclareSIUnit{\rcf}{rcf}
\usepackage[T1]{fontenc}
\renewcommand{\rmdefault}{qhv}
\usepackage{hyperref}

\begin{document}

\title[]{High-endurance mechanical switching in a DNA origami snap-through mechanism}

\abstract{Switchable elements are key components of dynamic technological and biological systems, enabling reversible transitions between well-defined states. Here, we present a DNA origami-based, mechanically bistable snap-through mechanism that can be electrically controlled. This nanoscale switch exhibits long-term stability in both states in the absence of external stimuli, while achieving millisecond-scale switching times upon application of an electric field. Individual devices sustain hundreds of thousands of switching cycles over several hours, offering a powerful platform for systematically studying the endurance and failure mechanisms of biomolecular nanoswitches. Functionalization with a gold nanorod further allows polarization-dependent optical modulation, opening avenues for applications in plasmonics. This versatile electromechanical interface has potential uses in molecular information processing, optical nanodevices, and the dynamic control of chemical reactions.}

\author[1,2]{\fnm{Florian} \sur{Rothfischer}} \email{florian.rothfischer@tum.de}
\author*[1]{\fnm{Lennart J. K.} \sur{Weiß}} \email{lennart.weiss@tum.de}
\author[1]{\fnm{Niccolò} \sur{Tedeschi}}
\author[1]{\fnm{Michael} \sur{Matthies}}
\author[1]{\fnm{Matthias} \sur{Vogt}}
\author[1]{\fnm{Christoph}\sur{Karfusehr}}
\author[1]{\fnm{Alexander} \sur{Hebel}}
\author[1,3]{\fnm{Petr} \sur{Šulc}}\email{psulc@asu.edu}
\author[2]{\fnm{Tim} \sur{Liedl}}\email{liedl@lmu.de}
\author*[1]{\fnm{Enzo} \sur{Kopperger}}\email{enzo.kopperger@tum.de}
\author*[1]{\fnm{Friedrich C.} \sur{Simmel}}\email{simmel@tum.de}
\affil[1]{\orgdiv{Department of Bioscience, TUM School of Natural Sciences}, \orgname{Technical University Munich}, \orgaddress{\street{Am Coulombwall 4a}, \city{Munich}, \postcode{D-85748}, \country{Germany}}}
\affil[2]{\orgdiv{Department of Physics}, \orgname{Ludwig Maximilians Universität München}, \orgaddress{\street{Geschwister-Scholl-Platz 1}, \city{Munich}, \postcode{D-80539}, \country{Germany}}}
\affil[3]{\orgdiv{School of Molecular Sciences and the Biodesign Institute}, \orgname{Arizona State University}, \orgaddress{\city{Tempe}, \postcode{85251}, \country{AZ,USA}}}
\maketitle

\section*{Introduction}
Switchable elements play a fundamental role in a wide range of natural and engineered systems, enabling transitions between distinct states in response to external stimuli \cite{howell2013compliant, cao2021bistable, chi2022bistable}. In technology, they form the basis of computation, data storage, and signal processing, whereas in biology, molecular switches regulate essential cellular processes, including gene expression, enzymatic activity, and signal transduction \cite{Cherry:2000km, Gardner:2000bm, xiong2003positive, ferrell2002self, Verdugo:2013cr}.

One particularly fascinating class of switches is that of mechanically bistable systems, which are employed as critical components in macroscopic engineering ranging from aerospace to medicine \cite{zirbel2016bistable, kota2005design}. These structures, such as snap-through mechanisms, remain stable in either of two configurations and transition between them only when a sufficient external force is applied. The design of nanoscale bistable switch equivalents to the ones used on the macroscopic scale presents a particular challenge: they must exhibit high stability in both states - in the presence of thermal noise - while remaining responsive to external control, allowing repeated transitions without compromising their integrity.

Recent advances in DNA nanotechnology have enabled the construction of dynamic nanoscale structures, including compliant mechanisms \cite{zhou2014dna} and electrically actuated DNA origami systems \cite{Pumm:2022aa,kopperger2018self}. However, the previous compliant designs have not yet demonstrated high stability, high-speed, and repeated switchability simultaneously~\cite{zhou2015direct, marras2015programmable}, and the switching transition was not directly monitored. In this work, we present a mechanically bistable DNA-based switch that achieves all of these features. Using external electric fields, we can reliably toggle the structure between its two states with unprecedented speed and endurance. This allows, for the first time, the study of device lifetime and failure mechanisms at the molecular scale, providing valuable insights for the development of robust nanomechanical systems. Our work demonstrates the potential of DNA-based bistable switches for applications ranging from nanophotonics to biochemical regulation.

\section*{Results} 
\subsection*{Design of a compliant bistable DNA origami structure}

\begin{figure*}[t]
\centering
\includegraphics[width=0.5\textwidth]{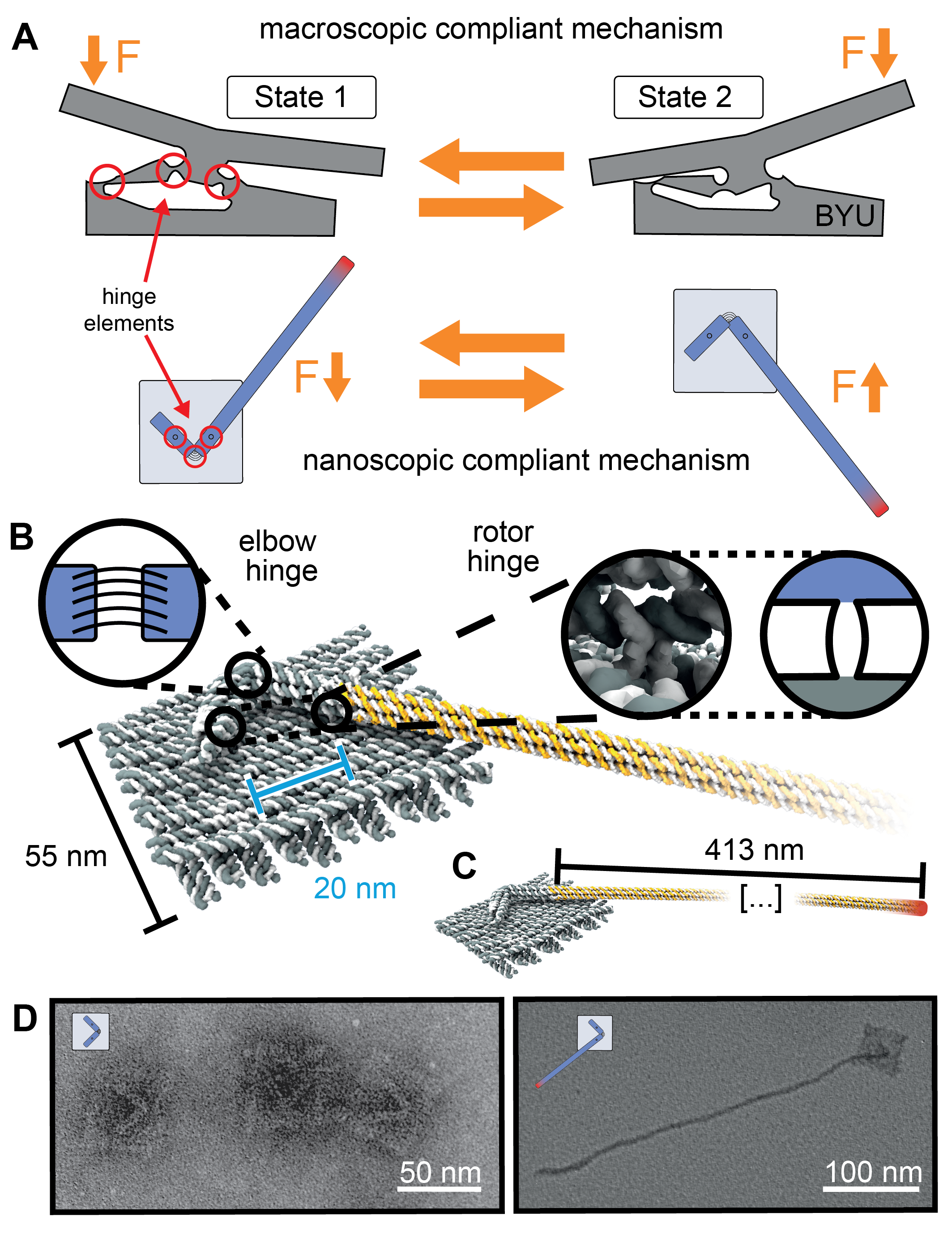}
\caption{{\bf Design of compliant bistable switch DNA-origami structure.}
(A) Exemplary macroscopic bistable compliant switch by Larry L. Howell group at Brigham Young University \cite{Jensen:2003} and implementation of a nanoscopic bistable switch. Yellow arrows show the needed points of force application to induce transition between states. Red circles marke the hinge positions. (B) DNA origami illustration of  oxDNA simulated structure of combined stator and bistable switch unit (grey) including the connection to the extended lever structure (yellow). The zoom-ins show the two types of hinge connections between the rotor units. Five staples connecting the rotor arms form the elbow hinge, while the rotor hinges are formed from single stranded scaffold between the stator and rotor units on both sides. (C) Full view of the construct including the fluorescent tip (red) on the extended lever structure, used for the tracking of switches. (D) TEM images of the bistable switch with and without attached extension lever structure. The insets shows the standard pictogram of the structure.}
\label{fig:Figure 1}
\end{figure*}

Compliant bistable mechanisms have two distinct mechanical equilibrium positions, which are stable in the absence of an externally applied force. Deformation of compliant elements allows to switch the structures between the two positions. A popular realization of mechanical bistability in macroscale engineering utilizes "snap-through-buckling" of a compliant structure, which features rigid parts connected by flexible hinge elements, to enable fast switching from one mechanical equilibrium to the other (cf. Figure~\ref{fig:Figure 1}A top) \cite{Jensen:2003}. 
Inspired by such mechanisms, we designed a DNA origami structure with two stable mechanical states, which are separated from each other by sterical hindrance (cf. Figure~\ref{fig:Figure 1}A bottom). 

Our structure consists of a pair of two six-helix bundle (6HB) rotor arms of \si{25} and \SI{26}{\nano\meter} length which are connected to the same \SI{55}{\nano\meter}~$\times$~\SI{55}{\nano\meter} base plate (cf. Figure~\ref{fig:Figure 1}B). Each of the two rotor arms is flexibly attached to the base plate via a rotor hinge, which is formed by \si{3}~and~\si{7} free nucleotides (nt) at the crossover site of the origami scaffold between base plate and rotor. The two rotor arms are linked to each other with a flexible elbow hinge composed of five staple strands, which in total generate six \si{15}~nt long oligo-dT flexible connections (cf. Figure~\ref{fig:Figure 1}B zoom panels).

Further, we attached an additional \SI{413}{\nano\meter} long 6HB extension arm to one of the short rotors (cf. Figure~\ref{fig:Figure 1}C), which we expected to act as a charged lever for electrical switching similar to our previous work.

Due to the geometric constraints set by the arm lengths and the hinge positions, in the relaxed state the elbow element is expected to make an angle of $\approx$~\SI{50}{\degree}, resulting in two possible relaxed configurations of the system (cf. Supplementary Note 1) .
We found that the origami structures indeed settle into one of these two stable mechanical states during folding (cf. Figure~\ref{fig:Figure 1}D).
Diffusive transitioning of the rotors between the states is inhibited by a sterical overlap of $\approx$~\SI{11}{\nano\meter}. However, we anticipated that the elbow structures might be compliant enough to allow switching from one of the mechanical states into the other upon application of an external force (Detailed origami design files and sequences, cf. Supplementary Note 1).

\subsection*{Electrical switching and single molecule data collection}

\begin{figure*}[t]
\centering
\includegraphics[width=1\textwidth]{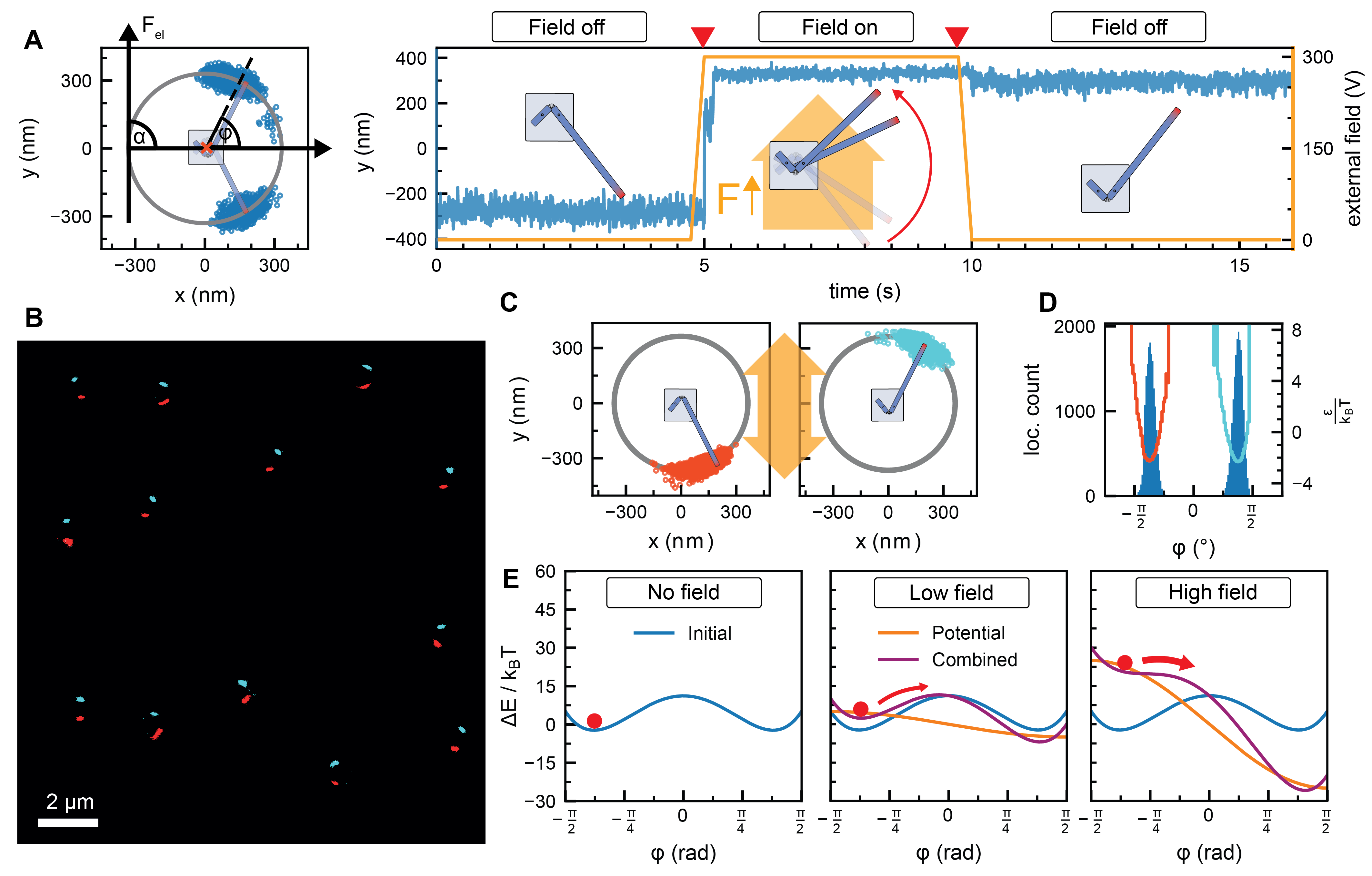}
\caption{{\bf Single switch experiments, experimental and theoretical observations of the structure.}
(A) Single switch particle localizations and time trace. The time trace shows the stability in the first state without applied field, the switching during the period where the field is applied, followed by staying in the second stable state after the field is turned off.
(B) Localization heat map of a single switch measurement. Red and cyan localizations show the two different stable states after a single switch. The image shows a stitched view of the total localization heat map of \si{13} exemplary structures, which successfully performed the single actuation. 
(C) Analysis of a single particle out of the measurement from panel \textit{B}, where the rotor was switched once but otherwise left without actuation for \SI{15}{\minute} before and after the induced switch.
(D) Determination of an energy landscape of the particle from panel \textit{C} based on Boltzmann inversion of the acquired location counts. The energy landscape shows two minima separated by a $\approx$~\SI{130}{\degree}. The calculated minima are $8-10 k_\text{B}T$ deep.
(E) Theoretical energy landscape based on fit of quartic potential to the data from \textit{D}. The three graphs show the landscape without applied field and under the influence of a low field and high external field input (field input = yellow, combined energy landscape = purple).
}
\label{fig:Figure 2}
\end{figure*}

For electrical switching experiments, we immobilized the bistable switch construct onto a functionalized glass substrate via six biotin-streptavidin linkages on the bottom of the stator plates. The mechanical state of each bistable switch was determined by tracking the fluorescently labeled tip of the extension arm, which carries \si{42}~fluorescent dyes, using single-molecule total internal reflection fluorescence microscopy (TIRFM). In the absence of an externally applied field, the extension undergoes Brownian movement in the neighborhood of one of the two expected mechanical equilibrium states, which is visualized by an accumulation of fluorescence localizations around the corresponding positions on the plane (cf. Figure~\ref{fig:Figure 2}A, left).

External electric fields for switching were applied using a computer-controlled actuation protocol, which was synchronized with the camera acquisition. Electrical switching of the bistable switches is observed by a transition of fluorescence localizations from one preferred position of the lever arm into the other. Importantly, the bistable switches stay in their new position also after removal of the electric field, demonstrating that the device is indeed bistable (cf. Figure~\ref{fig:Figure 2}A, right).

Switching efficiency depends on the orientation of the nanoswitches with respect to the external field. As our devices were randomly oriented on the glass substrate, we initially probed our ensembles with a voltage pulse of \SI{300}{\volt} for \SI{100}{\milli\second}. We then selected those structures for further analysis, which were successfully switched and were thus appropriately oriented with respect to the external field.
Figure~\ref{fig:Figure 2}B shows a heat map of the total localizations collected during a single switch experiment of similarly oriented switching structures. 
A circular fit of the fluorescent localizations allows to determine the relative position of the base plate, which is then used as the center of a local coordinate system. 
This allows extraction of the extension tip coordinates $x(t), y(t)$ and the angular position $\varphi(t)$ as a function of time.
Using a k-means clustering algorithm, the data is split into two clusters to determine the mechanical state of the structure at each time point, which allows us to calculate the number of successful switching cycles.

\subsection*{Energy landscape of the bistable switch}
When set into one of its two states, the pointer arm diffusively explores the neighborhood of the local energy minimum. Upon application of an external force, the arm can be switched into the other energy minimum, where, without applied external force, it again explores the local energy minimum (cf. Figure~\ref{fig:Figure 2}C).

From the distribution of the single-molecule localizations $p(\varphi)$ we can obtain the shape of the mechanical energy landscape $E(\varphi)$ close to the minima by Boltzmann inversion, i.e., $E(\varphi)=- k_B T \ln{p(\varphi)} + \mathrm{const}$ (cf. Figure~\ref{fig:Figure 2}D).
The minima are located at $\varphi \approx \pm 3\pi/8$ and are locally well approximated by parabolic potentials.

As desired, the minima are separated by a large energy barrier, which arises from the steric hindrance introduced by the overlap at the elbow hinge structure as well as the rigidity of the two 6HB rotors, which limits bending in the z-direction above the base plate to a certain degree. 
Due to the steepness of the barrier and the corresponding low number of data points in its neighborhood, the actual shape of the potential in the transition region cannot be reconstructed from the data, however. 

Several scenarios may explain how the bistable switches transition between states under applied force. One possibility is mechanical deformation of the connected 6HB rotor arms, resembling a macroscopic snap-through mechanism. Alternatively, force-induced melting of DNA duplexes in the origami structure may provide sufficient flexibility for the rotors to switch states. OxDNA simulations support a combination of all these aforementioned phenomena, implying that fraying helices near the elbow hinge together with bending and out-of-plane motion, enable the two 6HB rotor arms to pass one another (cf. Supplementary Note 2).

\subsection*{Comparison to theoretical estimations for the energy landscape}
Assuming a typical double-well potential $E(\varphi) \approx A\varphi^4 - B \varphi^2 + C$, we can estimate the height of the barrier and provide a semi-quantitative description of the dynamics of the bistable switch (cf. Figure~\ref{fig:Figure 2}E). A fit of the quartic potential to our data results in the parameters $A = 6.7~k_B T/\rm rad^4$, $B = 19~k_B T/\rm rad^2$, and $C = 11.2~k_B T$. This suggests an energy barrier of height $\Delta E = E(\varphi=0) - E (\varphi_{min})=B^2/4A \approx 13.6~k_B T$.

Due to the large barrier height we do not observe any thermally activated transitions, and in the absence of an externally applied force the system is trapped in one of its two minima.
We can estimate the lifetime of the state using the Kramers escape rate in the overdamped approximation \cite{Hanggi:1990en}, which can be expressed in terms of the parameters $A$ and $B$ as: $\sqrt{2} B \pi^{-1} \gamma_r^{-1} \cdot \exp{\left(-B^2/4 A k_B T\right)}$ (cf. Supplementary Note 3).


Here $\gamma_r$ is a frictional retardation term, which we assume to be on the same order as previously determined for a DNA origami rotor arm~\cite{vogt2023windup,rothfischer2024brownian}. With $\gamma_r = 1~\rm pN\cdot nm \cdot s \approx 0.24~k_B T \cdot s$, we obtain $k_{escape} \approx 4.5 \cdot 10^{-5}$~s$^{-1}$, or $\tau = 1/k_{escape} \approx 6~h$.
Our experiments showed that when monitoring \si{70} bistable switches in a single field of view for \SI{1}{\hour}, no spontaneous switching events were observed. This suggests that the lifetime of the switches may be even longer than this estimate.

In the presence of an externally applied electric field, the energy landscape is modified, resulting in 
$E(\varphi,V_0) \approx A\varphi^4 - B \varphi^2 + C + \xi V_0 \cos{(\varphi-\alpha})$, 
where $V_0$ is the bias voltage and $\alpha$ denotes the angular offset between the direction of the external field and the orientation of the bistable switch. 
$\xi$ is a coupling constant that converts between voltage and energy scale in our system, 
which we had previously determined to be $\xi \approx 0.1~k_B T/V$. 
Further shown schematically in the low and high field plot of Figure~\ref{fig:Figure 2}E, the external potential will modulate the energies of the minima and also the height of the barrier.
At high enough potentials, the barrier, depending on the offset $\alpha$, vanishes altogether (cf. Figure~\ref{fig:Figure 2}E high field plot).

\subsection*{Characterization of switching behavior}

\begin{figure*}[t]
\centering
\includegraphics[width=1\textwidth]{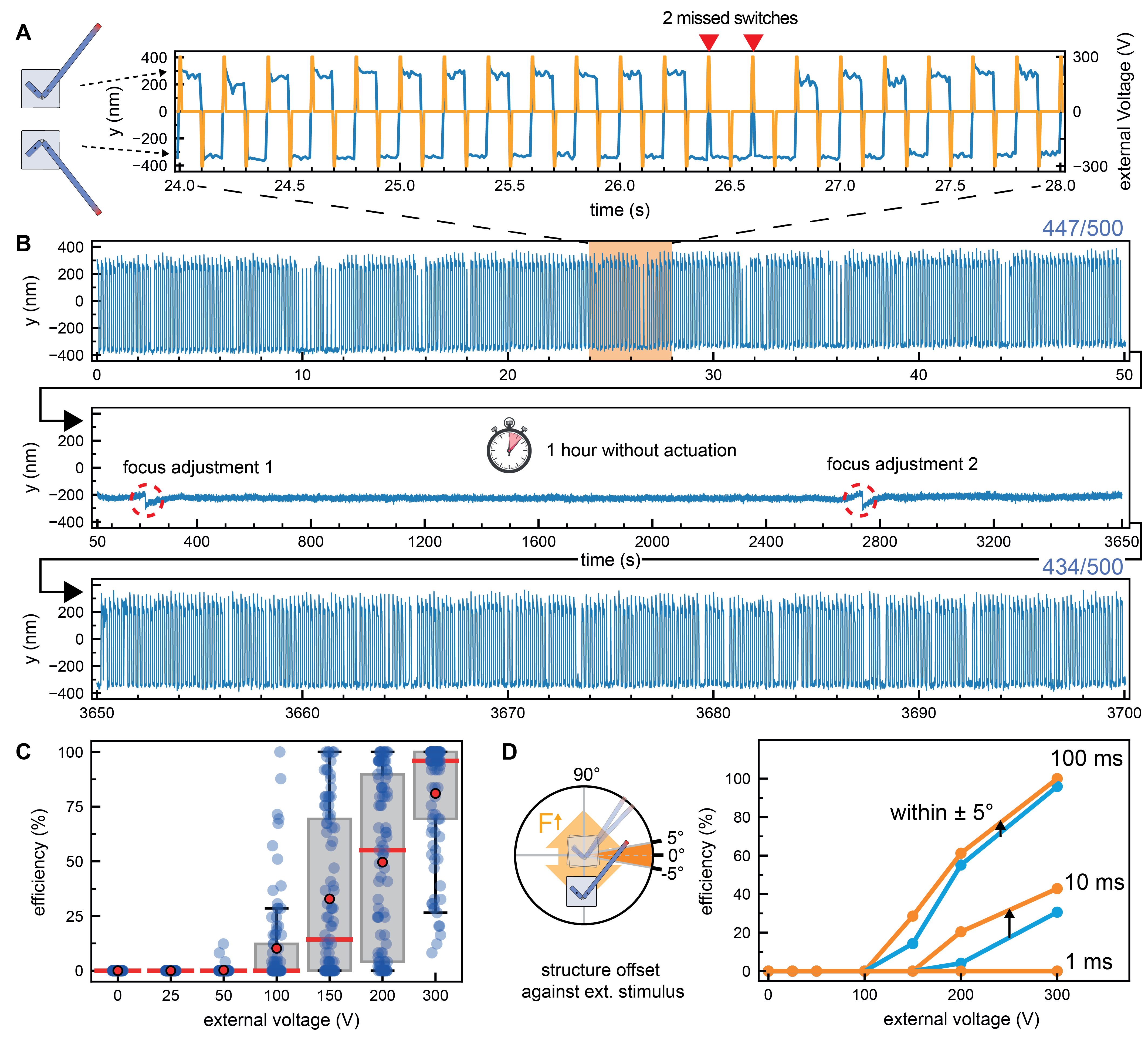}
\caption{{\bf Detailed analysis of the switching conditions.}
(A) Zoom in on the time trace shown in (B), visualizing switching over \SI{4}{\second}. The structure is actuated once per \SI{100}{\milli\second}. Red arrows show two missed switches.
(B) Single particle time trace. Three continuous phases: 1.) \si{500}~actuations over \SI{50}{\second}. 2.) \SI{1}{\hour} waiting period without external actuations. 3.) \si{500}~actuations over \SI{50}{\second}.
(C) Efficiency screen of the ensemble wide switching at voltages from \SI{0}{\volt} up to a maximum of \SI{300}{\volt} ($N= 93 particles$).
(D) Schematic visualizing the influence of axis offset against the applied external electric field. Efficiency comparison of the full ensemble (blue) against the ensemble low axis offset (brown) against the applied external electric field ($N=23 particles$) at three different pulse widths and six voltages.}
\label{fig:Figure 3}
\end{figure*}

We next assessed the performance of the bistable switch and its dependence on actuation time and electric field in greater detail.
When the electric field is switched slowly and with sufficient field strengths, appropriately oriented structures reliably switch from one state to the other.

However, for shorter pulse lengths, the structures sometimes fail to follow the electric field (cf. Figure~\ref{fig:Figure 3}A). We used the number of failed switching attempts to define a switching efficiency by dividing the number of successful mechanical switches by the number of the corresponding electric pulses. 

Figure~\ref{fig:Figure 3}B shows a time trace recorded from a bistable switch, which was electrically actuated \si{500}~times within \SI{50}{\second} (i.e. frequency of $f=\SI{10}{\hertz}$) with field pulses of \SI{10}{\milli\second} duration. Each field pulse was followed by a \SI{90}{\milli\second} pause to allow for heat dissipation occuring from the field application.
Upon this first switching period, the structure was left without actuation for a \SI{1}{\hour} interval, and then actuated with a second pulse run of \si{500}~actuations over \SI{50}{\second} (cf. Figure~\ref{fig:Figure 3}B second and third row plot).
The experiment shows that mechanical switching occurred in $447/500 =$~\SI{89}{\percent} and $434/500 =$~\SI{87}{\percent} of the cases under these conditions. Furthermore, the switch resided stably the last state of the first switching run for \SI{1}{\hour} without switching or disintegration, until it was further actuated.

We systematically changed the voltage to assess the dependence of the switching efficiency on the applied electrical field. While at low fields the energy landscape is not sufficiently tilted to induce transitions between the states, at high voltages of \SI{300}{\volt} (corresponding to a field strength of $E= 732~V/cm$) the bistable structure switches with a median efficiency of close to \SI{100}{\percent} (Figure~\ref{fig:Figure 3}C).

The response of the switches strongly depends on the alignment $\alpha$ of the structures with the externally imposed field.
In principle, the maximum torque is exerted when the field is perpendicular to the lever arm. However, in this case, after switching, the bistable system will be in a state from which it is difficult or even impossible to switch back when the field is reversed. The optimum field orientation for repeatedly switching the bistable switch back and forth turns out to be at an angle of $\alpha=\pi/2$, i.e., in the direction of the $y$-axis of the structure.

We thus filtered the ensemble of our voltage screen measurement from Figure~\ref{fig:Figure 3}C for particles with a maximal offset of their $x$-axis against the direction perpendicular to the applied external field of $\pm$~\SI{5}{\degree} (cf. Figure~\ref{fig:Figure 3}D rose plot). 
Comparison between these particles  with the full ensemble clearly shows higher switching efficiency, especially at lower switching voltages, for both \si{10}~and~\SI{100}{\milli\second} pulse length (Figure~\ref{fig:Figure 3}D plot).

This can be qualitatively understood in the following way: the applied voltage reduces the energy barrier for the transition, and thus the switching rate is higher at larger voltages. Only a fraction of the structures successfully switches within a given time window, but at \SI{100}{\milli\second} and at \SI{300}{\volt} the switching yield is almost \SI{100}{\percent}.
Further, for the optimum angle $\alpha=\pi/2$ the energy barrier for switching is the same in both directions and vanishes at $\approx$~\SI{240}{\volt}. At other angles correspondingly higher voltages are required to achieve the same switching efficiency. 
A semi-quantitative analysis of the switching yield with our rate model is given in Supplementary Note 3.

\subsection*{Plasmonic switching capabilities}

\begin{figure*}[t]
\centering
\includegraphics[width=1\textwidth]{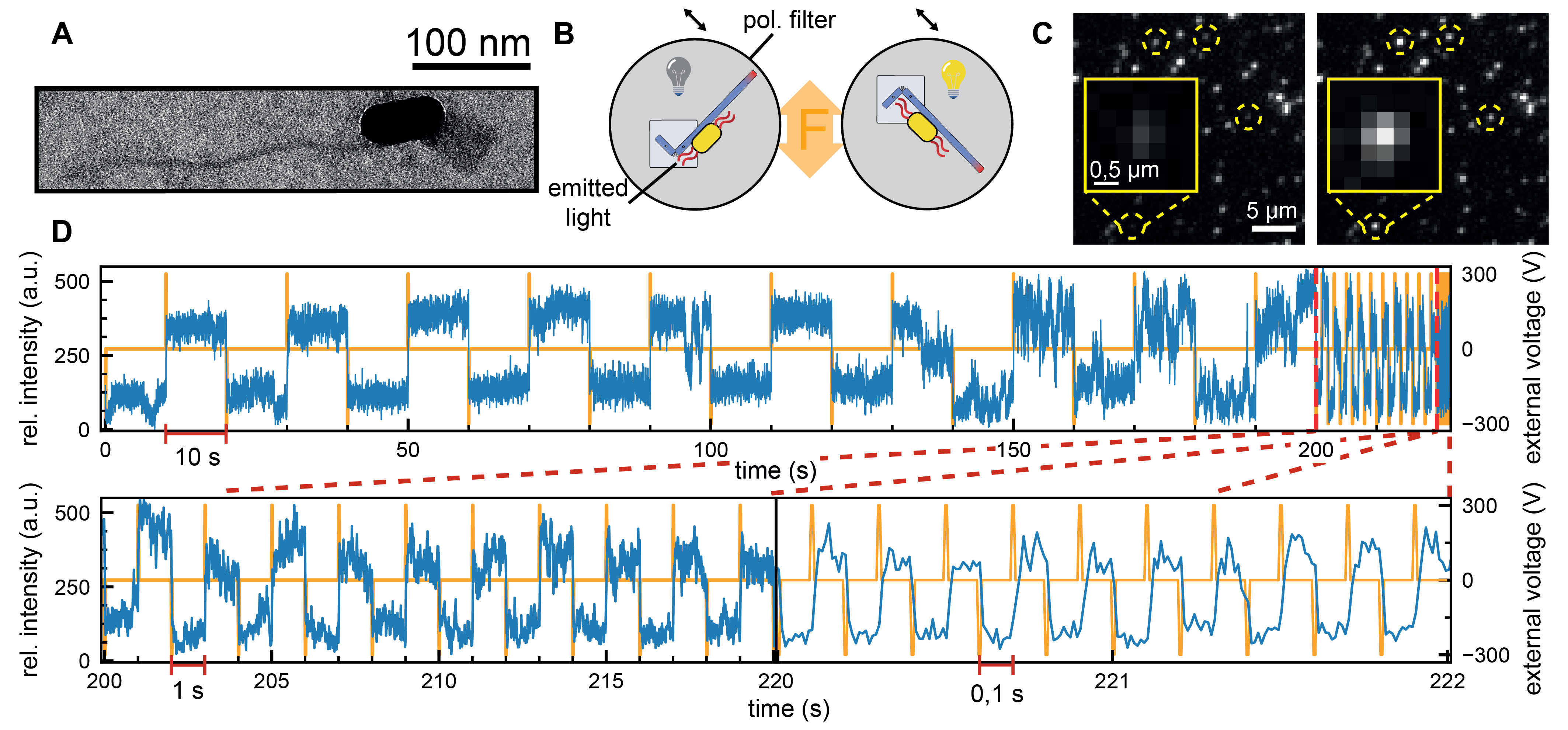}
\caption{{\bf A plasmonic toggle light switch.}
(A) TEM visualization of the bistable switch with an attached gold nanorod.
(B) Switching of the orientation of the gold nanorod can be monitored through a linear polarization filter in the detection path. When the rod is oriented perpendicular to the polarization axis of the detection filter, the observed scattering signal is maximal. 
(C) Raw microscopy data showing the two different states of four functional bistable structures (marked with yellow circles). The zoom-in shows the detailed signal change of a single particle.
(D) Time trace of plasmonic intensity switching over \SI{222}{\second}. The first \SI{200}{\second} the structure is switched every \SI{10}{\second} ($f=\SI{0.1}{\hertz}$). The two zoom-ins below show switching every \si{1}~($f=\SI{1}{\hertz}$)~and~\SI{0.1}{\second}~($f=\SI{10}{\hertz}$) over \si{20}~and~\SI{2}{\second} respectively.
}
\label{fig:Figure 4}
\end{figure*}

To demonstrate the potential of our bistable switch for applications in nanoplasmonics~\cite{Tan:2011by,Kuzyk:2012kc,kuzyk2014reconfigurable,Xin:2019kg} we modified the extension of the bistable switch with a gold nanorod with a width of \SI{25}{\nano\meter} and a length of \SI{60}{\nano\meter} (cf. Figure~\ref{fig:Figure 4}A). 

Excitation of the surface plasmon resonance in a gold nanorod results in a pronounced scattering signal. Nanorods typically exhibit two plasmon resonances, one along the long axis, the longitudinal mode, and a perpendicular one, the transversal mode. The scattered light of the longitudinal mode of a gold nanorod is linearly polarized along the nanorod’s long axis and appears for the given dimensions at $\lambda =$~\SI{650}{\nano\meter}. 
In our experiment, we thus illuminate the sample with unpolarized light at a wavelength of $\lambda =$~\SI{650}{\nano\meter}, and detect the scattered light through a linear polarization filter. The intensity of the measured scattering signal reaches its maximum when the gold nanorod is aligned in parallel to the polarization axis of the detection filter, and is nearly completely suppressed when the nanorod is oriented perpendicular to it (cf. Figure~\ref{fig:Figure 4}B).

Figure~\ref{fig:Figure 4}C shows representative raw data illustrating the intensity change upon switching of four functional bistable switches, as well as the detailed signal change of a single bistable switch.

To verify that plasmonic switching operates robustly and reliably, we actuated the gold nanorod-modified structure at frequencies of \SI{0.1}{\hertz}, \SI{1}{\hertz}, and \SI{10}{\hertz} (cf. Figure~\ref{fig:Figure 4}D). The time traces of the scattering signal demonstrate successful switching over \si{20}~cycles for all tested frequencies.

\subsection*{Maximal switching over 300k~actuations and fatigue analysis}


\begin{figure*}[t]
\centering
\includegraphics[width=1\textwidth]{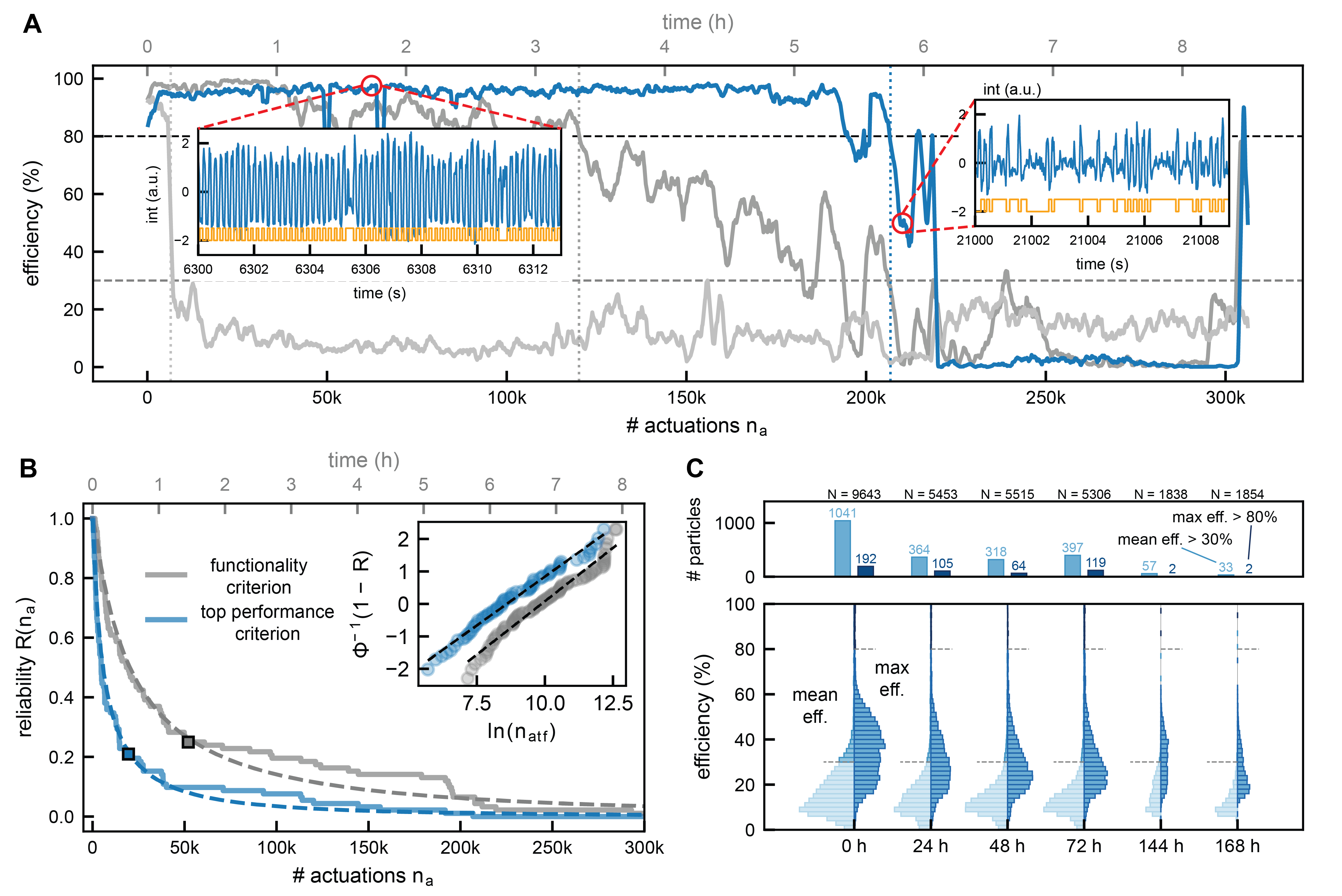}
\caption{{\bf Performance of the bistable switch over extended time spans.}
(A) Temporal evolution of switching efficiency for exemplary particles representing high (blue), medium (dark grey), and low (light grey) performance. Horizontal lines indicate the threshold for top performance and the minimum functionality cut-off. Vertical dashed lines mark the actuations-to-failure $n_{atf}$ based on the top performance criterion. Insets show intensity traces and state assignments before and after failure of a top performer.
(B) Device reliability based on actuations-to-failure for the top performance and functionality criteria. Experimental survival curves (solid) are shown alongside log-normal fits (dashed). Markers indicate the mean number of actuations to failure. The inset shows a log-normal probability plot.
(C) Switching efficiencies for the particle ensemble after resumed actuation at \si{0}, \si{24}, \si{48}, \si{72}, \si{144}, and \SI{168}{\hour}. Bottom: histograms show the mean (light blue) and maximum (dark blue) efficiencies of the ensemble during the first \SI{10}{\minute} of each day. Top: number of particles satisfying reaching mean efficiency \SI{>30}{\percent} and max efficiency \SI{>80}{\percent} at the respective time point. $N$ indicates the total number of remaining particles for the subsequent measurements.}
\label{fig:Figure 5}
\end{figure*}

One of the major advantages of a plasmonic readout is the absence of photobleaching. This makes it possible to monitor DNA nanodevices modified with metallic nanoparticles on the scale of several days instead of minutes, as possible with a fluorescent readout. 
We leveraged this capability to investigate an aspect of the bistable switch construct, which is normally hard to investigate with DNA nanodevices and that is particularly relevant for potential real-world applications: How often can such a device be actuated and in which way does repeated switching lead to fatigue and eventual failure? 

For the subsequent analysis we defined two classification criteria~\textemdash ~the top-performance criterion and the functionality criterion~\textemdash ~to distinguish gradual fatigue from outright failure. Under the top-performance criterion, bistable switches are considered as fatigued when their efficiency falls below \SI{80}{\percent} for three consecutive minutes, whereas under the functionality criterion, a switch is deemed non-functional if its efficiency remains below \SI{30}{\percent} for the same duration, indicating loss of responsiveness to external actuation.

To investigate the influence of excessive actuation, we actuated the bistable switches continuously for more than \SI{8}{\hour} and tracked their efficiency within \SI{20}{\second}-windows (cf. Figure~\ref{fig:Figure 5}A). 
The fatigue characteristics of our bistable switches were highly heterogeneous: i.e. one particle dropped out after \num{6500} actuations, whereas the best-performing particle fatigued after more than \SI{5.5}{\hour} and $206,800$ actuations, of which $195,000$ produced actual state transitions resulting in a mean efficiency of \SI{94}{\percent}.


Such a constant monitoring of actuations-to-failure $n_\mathrm{atf}$ allows us to calculate a reliability function (cf. Figure~\ref{fig:Figure 5}B) analogous to device failure modelling and maintenance prediction for macroscopic devices, where $R(n_\mathrm{a}) = P(n_\mathrm{atf} > n_\mathrm{a})$ is the probability that the bistable switch survives beyond $n_\mathrm{a}$ actuations. 
The two curves in Figure~\ref{fig:Figure 5}B indicate pronounced particle drop-out in the initial actuation phase, which could be due to misfolded structures. However, a subset of the particles do not show fatigue for a much higher number of actuations.

The robust fit of a long-tailed log-normal function, $R(n_\mathrm{a}) = 1 - \Phi\left( \frac{\ln(n_\mathrm{a}) - \mu}{\sigma} \right)$, where
$\Phi$ denotes the cumulative distribution function of the standard normal distribution, to our data indicates that the ensemble of switches is very heterogeneous and that a few devices perform significantly better than the majority of the population (cf. Figure~\ref{fig:Figure 5}B inset).
Using the fitted distribution, we obtain a median number of operations to failure of $5,682$ for the top performance-criterion and $20,254$ for the functionality criterion. The mean number of actuations before drop-out are found to be $19,681$ and $61,011$, respectively (cf. square marker Figure~\ref{fig:Figure 5}B). 




The distribution further suggests that failure is likely driven by multiplicative processes, resulting from the combined effects of stochastic, device-intrinsic factors and independent external influences. A variety of external conditions may affect switching during the experiment— i.e. diffusing contaminants may obstruct the switch, the substrate surface may degrade, or buffer conditions may change over due to electrochemical reactions at the electrodes.
Notably, however, unlike typical macroscopic devices, some bistable switches in our ensemble exhibit partial recovery of switching behavior or begin switching only after a delay. This observed `healing' effect may arise from factors similar to those discussed for drop-out events, and may additionally involve staple rebinding and reflect inherent resilience or self-repair capabilities of the DNA origami components.

To assess long-term degradation in switching performance, we resumed actuation of the same bistable switch sample (Figure~\ref{fig:Figure 5}A,B) after 1, 2, 3, 6, and 7 days. For each time point, we analyzed the first \SI{10}{\minute} of switching using a mean and max efficiency criterion. We then quantified the efficiencies of the remaining active particles (Figure~\ref{fig:Figure 5}C violin plots), and determined how many particles still performed according to a \SI{30}{\percent} mean- or \SI{80}{\percent} max-efficiency criterion (cf. Figure~\ref{fig:Figure 5}C bar plot). 
Notably, \SI{24}{\hour} after the initial measurement and \SI{8}{\hour} out of those continously actuated, $364$ particles still exceeded the \SI{>30}{\percent} mean-efficiency threshold, and $105$ surpassed the \SI{>80}{\percent} maximum-efficiency threshold. By day \si{7}~(\SI{168}{\hour}), these numbers had declined, but $33$ and $2$ particles respectively still showed switching activity.

\section*{Discussion}
In conclusion, we developed a compliant, bistable DNA origami switch that can be actuated with high precision and reproducibility over multiple switching cycles using externally applied electric fields.
By applying principles of compliant mechanism design borrowed from macroscale systems, we distributed mechanical strain across multiple structural elements, thereby increasing the device’s durability and enabling stable, repeatable switching behavior.
We characterized the switching dynamics in detail using single-molecule fluorescence tracking and showed that switching efficiency depends on both the applied torque (via the electric field) and the actuation duration.
From the single-molecule trajectories, we derived an effective energy landscape of the device, which allowed us to rationalize the observed switching behavior using a simplified theoretical model.
 
An important feature of our bistable switch is that the external electric field only needs to be applied for a millisecond time duration to induce a stable state transition. In the absence of the electric field, the system remains stably locked in one of its two alternative states despite the Brownian fluctuations that are predominant at the nanoscale. This behavior contrasts with previously developed electromechanical DNA origami devices, which had to be held in place either by a continuous electric field or through hybridization to the substrate via DNA linkers~\cite{kopperger2018self,vogt2023windup}.

We exploited the enhanced switching performance and stability of our system by implementing a plasmonic switching device: gold nanorods were attached to the bistable switch, enabling electrical control of the polarization-dependent plasmonic scattering signal. This signal also provided a non-bleaching, single-molecule readout.
Using this approach, we were able to systematically investigate device stability and fatigue for the first time in a DNA origami nanodevice. In particular, we showed that individual switches can undergo hundreds of thousands of reversible transitions over several days - highlighting the potential of our system for real-world applications such as biosensing or nanoscale displays.

Our development of a bistable, compliant DNA origami switch opens up a broad range of potential future directions. On a fundamental level, the switch can serve as a basic information storage unit, capable of encoding one bit~\cite{Doricchi:2022aa,Wang:2023}. Arrays of such electromechanical switches could be coupled to create more complex switching networks capable of performing logic operations and information processing tasks.
Integration with nanoscale fabrication techniques - such as electron-beam lithography for precise positioning and alignment~\cite{Scheible:2014gv,Gopinath:2016ea,Gopinath:2021} - could enable the construction of hierarchical actuation lines or circuits, allowing synchronized switching across larger assemblies.

A particularly exciting avenue is the use of these switches as electromechanical interfaces for controlling biochemical reactions, potentially providing a long-sought bridge between electronics and biology. For example, one could envision devices in which active sites - such as enzyme binding pockets, substrates, or substrate docking sites - are sterically hidden in one of the switch states and exposed in the other, thereby enabling or disabling downstream biochemical processes \cite{rosier2020proximity}.

In this context, the fact that the electric field only needs to be applied transiently during the switching event, and can remain off otherwise, is especially advantageous. It avoids side effects associated with continuous electric fields, such as undesired electrochemical reactions, which often complicate conventional bioelectronic approaches.

\section*{Methods}

\subsection*{DNA origami design}
Structures were designed using cadnano~2.5 (cf. reference~\cite{douglas2009cadnano}). The stator structures including the two small rotors was folded using a $8,064$~bases scaffold. The extension lever structure was folded using a $7,249$~long scaffold. Detailed cadnano designs and sequences can be found in Supplementary Note 1.

\subsection*{Folding}
Folding of DNA origami samples was typically performed by using \SI{50}{\micro\liter} of \SI{100}{\nano\molar} scaffold (in-house produced in dd\ce{H2O}) combined with a $5~\times$ molar excess of staples (obtained from IDT at \SI{100}{\micro\molar} in IDTE buffer) over the scaffold. For the stator structure we adjusted the sample buffer to 1xTE, \SI{20}{\milli\molar}~\ce{MgCl2} and \SI{5}{\milli\molar}~\ce{NaCl}, while we used lower \SI{12}{\milli\molar}~\ce{MgCl2} version for the extension lever structure as folding buffer. Samples were thermally annealed using a thermal cycling device. The thermal annealing ramp was performed from \si{70} to \SI{40}{\celsius}, reducing the temperature by \SI{0.1}{\celsius} every \SI{2}{\minute}. 

\subsection*{Purification and assembly}
Purification was performed in step-wise manner. First, using PEG-precipitation was done similar to reference \cite{stahl2014facile}. After the PEG precipitation we added Neutravidin in $10~\times$~molar excess over the initial scaffold concentration of the sample during folding for the stator structure. For the extension lever structure we added fluorescent dyes, with a complementary sequence, in $5\times$~excess over the total number of marker extensions. As a last purification step we performed agarose gel electrophoresis (\SI{70}{\volt}, \SI{1}{hour}, $0.5~\times~TBE$, \SI{12}{\milli\molar}~\ce{MgCl2} running buffer) followed by gel extraction similar to reference \cite{rothfischer2024brownian}. The stator structure was extracted based on fluorescence of an Atto488 marker dye added during folding, while the extension lever was extracted through the fluorescence from the added Atto655 dyes. The stator and extension lever were combined at a $1:1$ ratio for the final assembly and incubated at \SI{37}{\celsius} for \SI{30}{\minute}.

\subsection*{Transmission electron microscopy}
Negative stain transmission electron microscopy micrographs were recorded as pusblished previously \cite{buechl2022energy} using a Philips CM-100 at $100~kV$ and FCF400-CU copper grids from Electron Microscopy Sciences.

\subsection*{Sample preparation for TIRF microscopy}
The objective-based TIRF setup is the same as used in reference \cite{vogt2023windup}. Fully assembled structures were applied to a PEG-modified glass slide (modification protocol cf. reference~\cite{kopperger2018self}) at a concentration of \SI{200}{\pico\molar} with a measurement buffer of $0.3~\times~TB$, \SI{3}{\milli\molar}~\ce{MgCl2}, \SI{100}{\milli\molar}~\ce{NaCl}. For the measurement in Figure~\ref{fig:Figure 2}A the \SI{100}{\milli\molar}~NaCl was omitted. Sample chambers used for fluorescence measurements as well as the electrodes used for field application are the same as used in reference~\cite{vogt2023windup}. After the sample was applied to the PEG-biotin surface, the chamber was washed with at least $5~\times$ total volumes of the measurement chamber. For the measurement, the chamber and wells were fully filled with measurement buffer. The electrical actuation setup was the same as in \cite{vogt2023windup}.

\subsection*{Analysis of single molecule fluorescence measurements}
Recorded videos of switching routines using fluorescence tracking were processed using the \textit{Picasso} software package \cite{schnitzbauer2017paint}. The combined emission of the ~\si{42} fluorescent dyes at the tip of the extension structure were treated as a single emitter and used for the positional localization of the bistable switching structure. If necessary, the drift correction functions of \textit{Picasso} were applied to the data set using the \textit{localize} or \textit{render} package. Particles that showed the expected localization and movement pattern were picked manually and their event list was exported for subsequent analysis using a custom python-based routine. 

As a first step in the python analysis routine, the center of each particle, which also represents the center of the bistable structure, was determined. This was achieved using a circular fit. In a next step, the positional coordinates of all localizations were transformed into relative polar coordinates per particle. Particle trajectories were generated by plotting the normalized particle position as function of average time. 

To calculate the switching efficiency the data was fed into a k-means clustering algorithm to learn and predict the two particle states. Based on the predicted state sequence, the number of switching events per particle was quantified. To reduce false-positive classifications, a minimum dwell time threshold was applied to each state. False-positive switches typically arose in particles whose lever extensions followed the electric field actuation but rapidly returned to their baseline state once the external trigger ceased.

The vector connecting the centers of the two state clusters was used to calculate the offset angle between the observed lever response and the direction of the externally applied electric field. The opening angle of the motion was then determined by relating this particle vector to the center of the fitted circular trajectory.

The energy landscapes were calculated from the angular data and converted via Boltzmann inversion $U(x) = -k_B T \ln P(x) + C$. Bins without any localizations $P(x)$ were manually set to a value of $10~k_B T$ to ensure the calculated energy remains finite.

\subsection*{Sample preparation for plasmonic scattering measurements}
Gold nanorods were prepared following the low pH route according to \cite{kuzyk2014reconfigurable}.The DNA origami structures and the sample chamber were prepared as previously described, with the additional step of incubating gold nanorods onto the surface-bound origami structures. The gold nanorods were adjusted to \SI{10}{\nano\molar} in $1~\times$ measurement buffer before \SI{20}{\micro\liter} were added to the microscopy sample chamber and incubated with the origami structures at RT for \SI{30}{\minute}. Unbound gold nanorods were washed off using $1~\times$ measurement buffer.
For the extended actuation experiments, a single channel containing bistable switches was prepared and continuously measured over seven days while remaining on the microscope stage. During external actuation, the buffer was exchanged every hour, and the channel was sealed during overnight resting periods (cf. Figure~\ref{fig:Figure 5}). 

\subsection*{Scattering microscopy setup}
 Scattering microscopy was used for samples modified with gold nanorods. The system consisted of an inverted optical microscope (Olympus IX71) which allowed for the illumination of the sample with obliquely incident light. A high-intensity red light source (\SI{3}{\watt}~\SI{650}{\nano\meter} LED, Winger Electronics GmbH and Co. KG) was used for slide waveguide-based illumination, and the scattered light was collected using an $40~\times$~objective lens with a numerical aperture (NA) of $0.75$ (Olympus UPlanFI). Images were recorded using a high-sensitivity CMOS camera (ORCA-Fusion Digital CMOS camera C14440, Hamamatsu Photonics, Japan) with a typical exposure time of \SI{20}{\milli\second} (\SI{50}{fps}). For polarization-dependent measurements a linear polarization filter was introduced into the emission beam path. Measurements were synchronized with the externally applied field as in previous experiments. Light scattering data were acquired with a frame time of \SI{20}{\milli\second}, while bistable switches were actuated using a \SI{300}{\volt}, \SI{10}{Hz} signal with a pulse width of \SI{10}{\milli\second}.

\subsection*{Analysis of scattering microscopy data}
Initial particle positions and particle drifts were determined in a custom Python script using functions from \textit{Picasso}'s \textit{localize} and \textit{render} packages to extract intensity values per frame and per particle \cite{schnitzbauer2017paint}. The resulting intensity traces were low-pass filtered and standardized via z-score normalization. Change point detection was applied to segment the data into windows of similar signal envelopes, which were subsequently analyzed using a two-state hidden Markov model (HMM) classifier to infer probable state transitions. A decision tree, based on signal amplitude, coherence with a biphasic reference signal, and the power spectral densities at \SI{5}{\hertz} and \SI{10}{\hertz}, was used to categorize the windows and optimize HMM learning and prediction. State sequences were post-processed, retaining only state transitions that satisfied a minimum dwell time of three frames per state and exhibited well-separated intensity differences.

The reliability plot shown in Figure~\ref{fig:Figure 5}B for the \SI{8}{\hour} actuation experiment was generated from an initial ensemble of well-behaving particles, selected based on exceeding \SI{80}{\percent} switching efficiency at least once within the first \SI{3}{\minute}. Experimental time-to-failure was determined based on two criteria: a decline below \SI{80}{\percent} (top performance) or below \SI{30}{\percent} (functionality) for three consecutive minutes. The particles' mean and maximum switching efficiencies in Figure~\ref{fig:Figure 5}C were calculated from the initial \SI{10}{\minute} of each measurement day.

\section*{Data Availability}
Source data and all other data that support the plots within this paper and other findings of this study are available from the corresponding authors upon reasonable request.

\section*{Code Availability}
The source code of the data analysis routines and simulation files employed in this study is available from the corresponding authors upon reasonable request.

\section*{Acknowledgments}
We thank the group of H. Dietz for providing in-house produced scaffold strands. We thank M. Dass and G. Posnjak for the introduction to the modification of gold nanoparticles and dark-field scattering microscopy. This work was funded by the Deutsche Forschungsgemeinschaft (DFG, German Research Foundation) through SFB1032 (project ID 201269156 TPA2 and TPA6). We gratefully acknowledge funding by the BMBF through its 6G-life initiative. We acknowledge funding by the BMBF and the Free State of Bavaria under the Excellence Strategy of the Federal Government and the Länder through the ONE MUNICH Project Munich Multiscale Biofabrication. P.Š. acknowledges funding from the ERC under the European Union’s Horizon 2020 research and innovation program (grant agreement number 101040035)). We acknowledge funding for A.H. by the European Innovation Council EIC (project MI-DNA DISC, Grant agreement ID: 101115215). C.K. gratefully acknowledges funding through the Max-Planck-School Matter to Life, which is supported by the German Federal Ministry of Education and Research (BMBF) in collaboration with the Max Planck Society. F.R. has personally been funded by the Konrad-Adenauer-Stiftung (KAS). 

\section*{Author Contributions}
F.R. and E.K. designed the DNA nanostructures. F.R. produced and purified the DNA nanostructures. F.R. produced the modified nanoparticles. F.R. performed the single molecule measurements. E.K. and L.W. gave advice on recorded data. L.W., A.H. and C.K. recorded the TEM images. L.W., F.R. and M.V. programmed the data analysis routines. F.R. performed the oxDNA simulations used for structure rendering. N.T., M.M., and P.S. performed oxDNA simulations of the switching process. F.R., E.K., L.W., T.L., and F.C.S. planned the research. F.C.S. and F.R.  developed the theoretical model. T.L. and F.C.S. acquired the funding. All authors discussed the scientific results and contributed to the writing process.

\section*{Competing interests}
F.C.S, F.R. and E.K. have filed a patent application related to the technology discussed in this manuscript (Application number EP4550985A1).

\section*{Additional information} 
Supplementary information is available at XX. Correspondence and requests for materials should be addressed to Lennart Weiß, Enzo Kopperger or Friedrich C. Simmel.

\bibliography{scibib}
\end{document}


\graphicspath{{Figures/Supplement}}

\begin{titlepage}
\centering
\vspace*{2cm}

{\LARGE\bfseries
High-endurance mechanical switching in a DNA origami snap-through mechanism\\[0.5cm]
Supplementary Information
\par}

\vspace{1.5cm}

Florian Rothfischer\textsuperscript{1,2},
Lennart Weiß\textsuperscript{1,*},
Niccolò Tedeschi\textsuperscript{1},
Michael Matthies\textsuperscript{1},
Matthias Vogt\textsuperscript{1},
Christoph Karfusehr\textsuperscript{1},
Alexander Hebel\textsuperscript{1},
Petr Šulc\textsuperscript{1,3},
Tim Liedl\textsuperscript{2},
Enzo Kopperger\textsuperscript{1,*},
Friedrich C. Simmel\textsuperscript{1,*}

\vspace{1.5cm}

\textsuperscript{1}Department of Bioscience, TUM School of Natural Sciences, Technical University Munich, D-85748 Garching, Germany\\
\textsuperscript{2}Department of Physics, Ludwig Maximilians Universität München, Geschwister-Scholl-Platz 1, D-80539 Munich, Germany\\
\textsuperscript{3}School of Molecular Sciences and the Biodesign Institute, Arizona State University, Tempe, AZ 85251, USA

\vspace{1cm}

*Corresponding authors: lennart.weiss@tum.de, enzo.kopperger@tum.de, simmel@tum.de

\vfill
\end{titlepage}

\newpage

\tableofcontents

\newpage

\section{DNA origami design details}

\subsection*{Dimensions and design files}
\begin{figure*}[htbp!]
\centering
\includegraphics[width=1\textwidth]{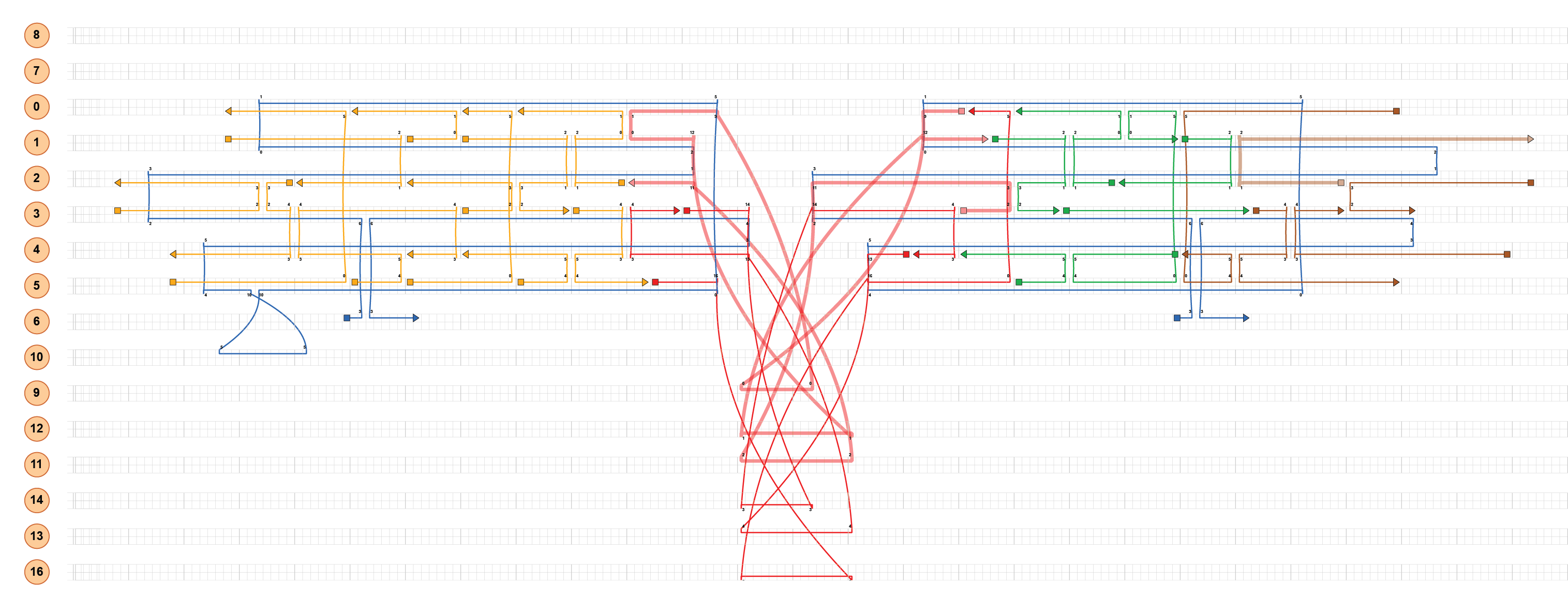}
\caption{\textit{caDNAno} design of the two connected rotor elements. The two rotor elements are connected by five staples with a \SI{15}{nt} long Poly-T spacer.}
\label{SI3}
\end{figure*}

\begin{figure*}[htbp!]
\centering
\includegraphics[width=1\textwidth]{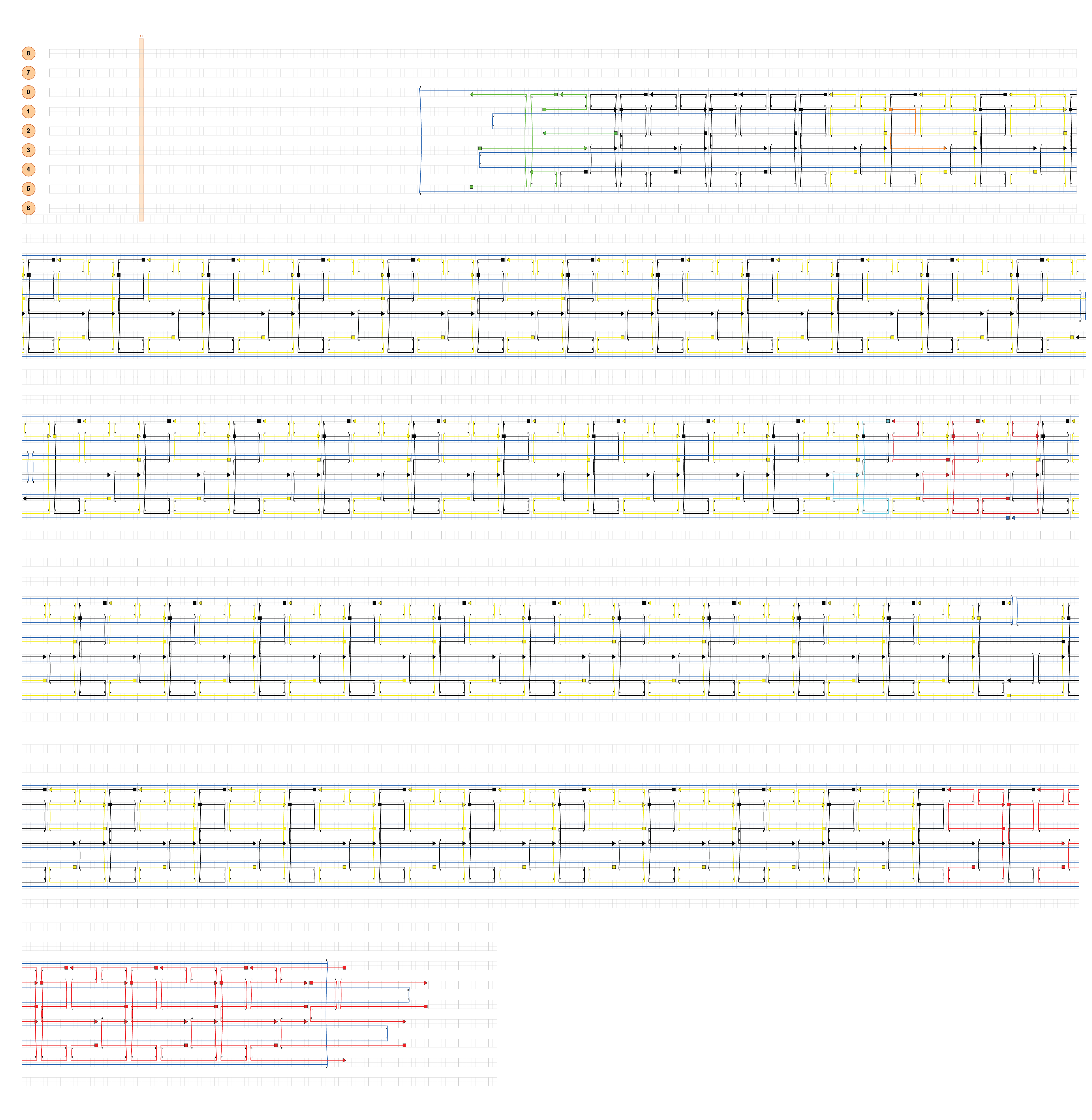}
\caption{\textit{caDNAno} design of the extension lever structure without modification. The structure has the rotor binding site on the left and the fluorescent marker modification on the right tail end.}
\label{SI4}
\end{figure*}

\begin{figure*}[htbp!]
\centering
\includegraphics[width=1\textwidth]{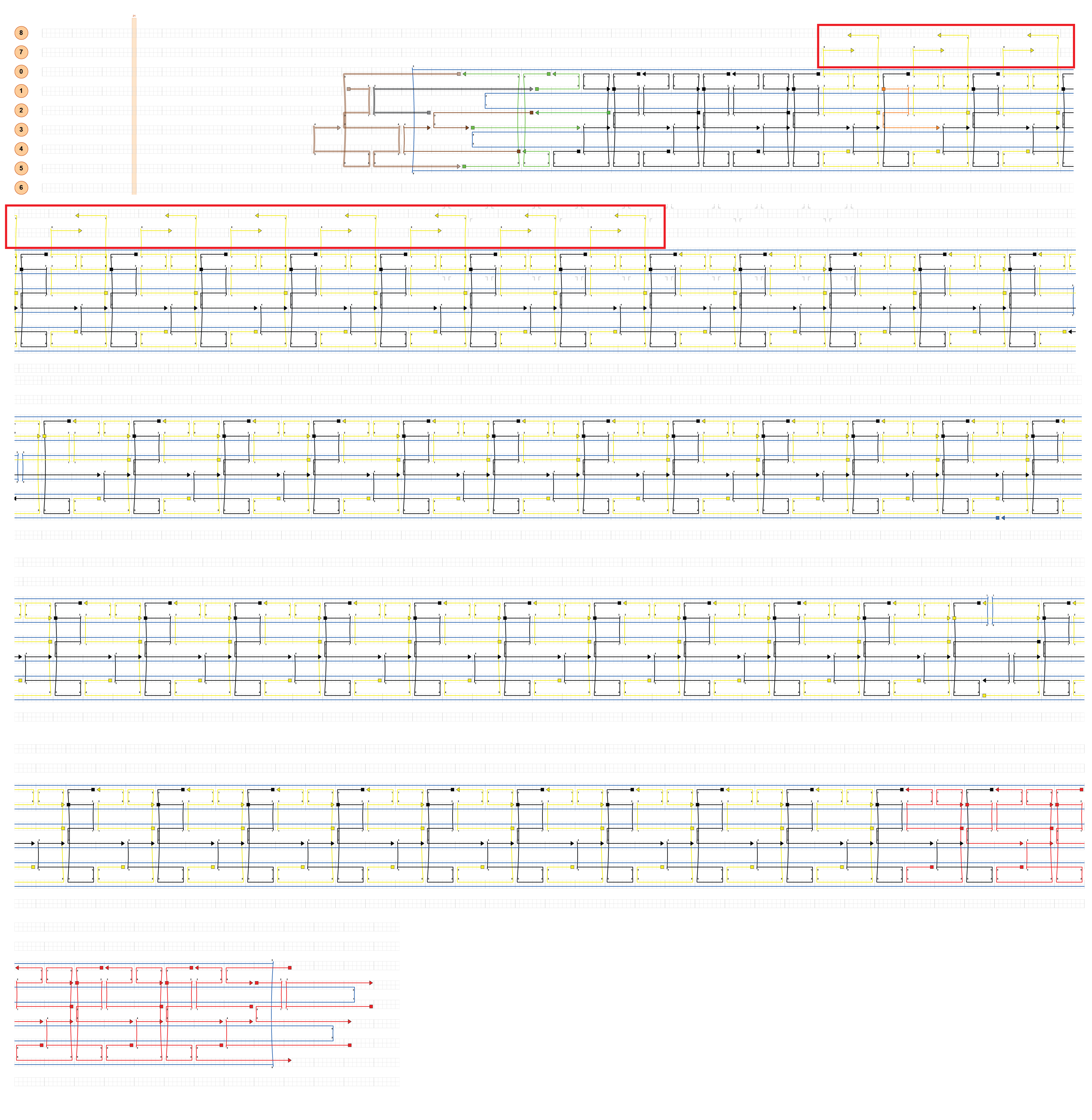}
\caption{\textit{caDNAno} design of the extension lever structure with binding sites for a Gold Nanorod. The extension lever scaffold and staple routing is identical to the one shown in Supplemental Figure~\ref{SI4}, but \si{16} staples close to the binding site for the rotor elements are elongated with a \SI{8}{nt} Poly-A tag for the Gold Nanorod binding (red box).}
\label{SI5}
\end{figure*}

The full bistable switch structure consists of three design parts, which are folded in two separate folding reactions. Key dimensional parameters of the structure are shown in Supplemental Figure~\ref{SI1}. The design shown in Supplemental Figure~\ref{SI2} contains the design of the stator structure, while Supplemental figures~\ref{SI3} and \ref{SI4}/\ref{SI5} contain the two rotor six helix bundles (6HBs) of the bistable switch mounted on top of the stator unit as well as the extension lever, with and without binding sites for the gold nanorod, attached to the bistable switch and stator construct. 

The stator is a double-plate DNA origami made from two crossed-square sheets of \SI{55}{\nano\meter} side length using the first \si{7,174}~bases of the \si{8,064}~base scaffold (as published previously in \cite{kopperger2018self, vogt2023windup}). The bistable switch is made up from two 6HB structures, which are interconnected to each other by five staples. All of the five connecting staples contain a \si{15x}~Thymine nucleotide spacer between the two binding parts to allow for and extended mobility of the two 6HB structures. Further the bistable switch uses the same scaffold as the stator unit using the last \si{890}~bases (including \si{62} bases connecting the two rotors on helix \si{11} of the stator from \si{11[135]} to \si{11[75]}) of the \si{8,064}~scaffold.

Separately from the stator and bistable switch unit, the extension lever is folded as a 6HB bundle. This 6HB bundle is connected to one of the bistable switch unit 6HB bundles after folding and purification. For further details, compare designs to our previous publications from Kopperger et al. \cite{kopperger2018self}, Vogt et al. \cite{vogt2023windup} and Rothfischer et al. \cite{rothfischer2024brownian}.

\begin{figure}[htbp!]
\centering
\includegraphics[width=0.5\textwidth]{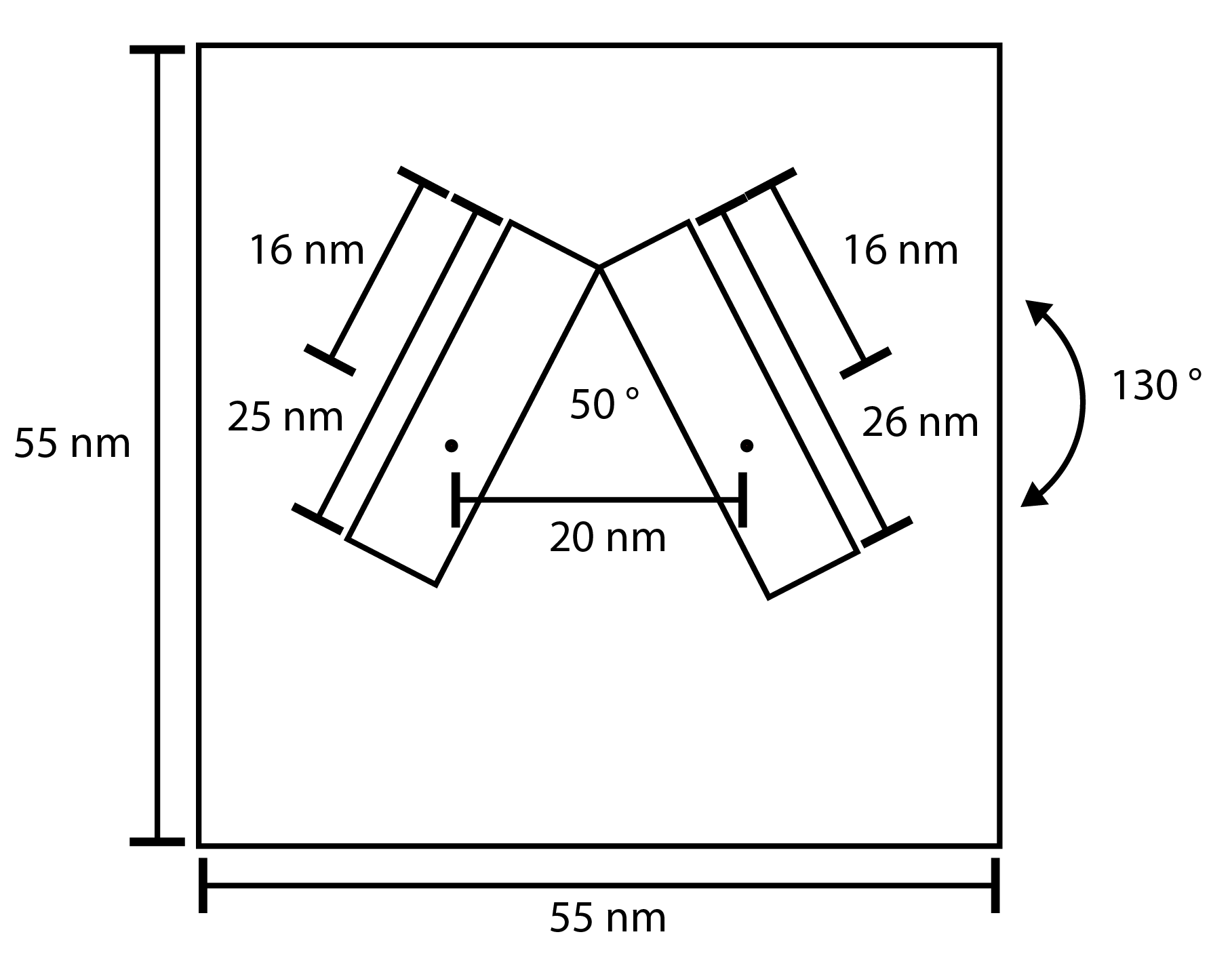}
\caption{Sketch of the bistable switch with dimensional annotation of the key parameters.}
\label{SI1}
\end{figure}

\begin{figure*}[htbp!]
\centering
\includegraphics[width=1\textwidth]{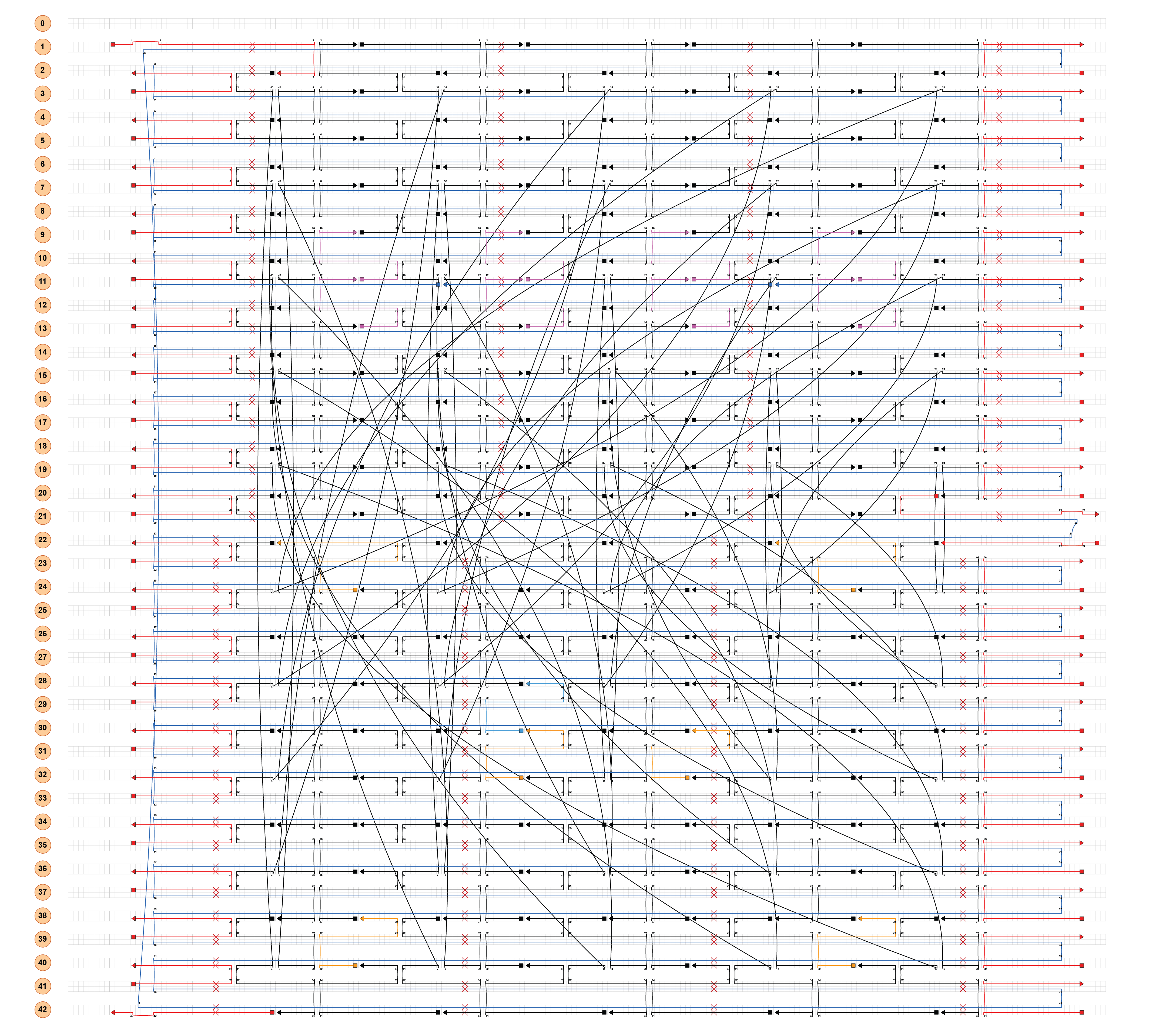}
\caption{\textit{caDNAno} design of the stator construct. The figure illustrates the routing of scaffold and staple strands. The stator consists of two plates stacked in a crossed conformation. Staples marked in orange indicate strands modified with biotin.}
\label{SI2}
\end{figure*}

\clearpage
\subsection*{Scaffold versions}

\subsubsection*{8,064 scaffold}
\si{8,064}~basepair scaffold version used for the bistable switch structure:
\begin{verbatim}
1         11        21        31        41        51        61        71
AAGCTTGGCACTGGCCGTCGTTTTACAACGTCGTGACTGGGAAAACCCTGGCGTTACCCAACTTAATCGCCTTGCAGCAC
81        91        101       111       121       131       141       151
ATCCCCCTTTCGCCAGCTGGCGTAATAGCGAAGAGGCCCGCACCGATCGCCCTTCCCAACAGTTGCGCAGCCTGAATGGC
161       171       181       191       201       211       221       231
GAATGGCGCTTTGCCTGGTTTCCGGCACCAGAAGCGGTGCCGGAAAGCTGGCTGGAGTGCGATCTTCCTGAGGCCGATAC
241       251       261       271       281       291       301       311
TGTCGTCGTCCCCTCAAACTGGCAGATGCACGGTTACGATGCGCCCATCTACACCAACGTGACCTATCCCATTACGGTCA
321       331       341       351       361       371       381       391
ATCCGCCGTTTGTTCCCACGGAGAATCCGACGGGTTGTTACTCGCTCACATTTAATGTTGATGAAAGCTGGCTACAGGAA
401       411       421       431       441       451       461       471
GGCCAGACGCGAATTATTTTTGATGGCGTTCCTATTGGTTAAAAAATGAGCTGATTTAACAAAAATTTAATGCGAATTTT
481       491       501       511       521       531       541       551
AACAAAATATTAACGTTTACAATTTAAATATTTGCTTATACAATCTTCCTGTTTTTGGGGCTTTTCTGATTATCAACCGG
561       571       581       591       601       611       621       631
GGTACATATGATTGACATGCTAGTTTTACGATTACCGTTCATCGATTCTCTTGTTTGCTCCAGACTCTCAGGCAATGACC
641       651       661       671       681       691       701       711
TGATAGCCTTTGTAGATCTCTCAAAAATAGCTACCCTCTCCGGCATTAATTTATCAGCTAGAACGGTTGAATATCATATT
721       731       741       751       761       771       781       791
GATGGTGATTTGACTGTCTCCGGCCTTTCTCACCCTTTTGAATCTTTACCTACACATTACTCAGGCATTGCATTTAAAAT
801       811       821       831       841       851       861       871
ATATGAGGGTTCTAAAAATTTTTATCCTTGCGTTGAAATAAAGGCTTCTCCCGCAAAAGTATTACAGGGTCATAATGTTT
881       891       901       911       921       931       941       951
TTGGTACAACCGATTTAGCTTTATGCTCTGAGGCTTTATTGCTTAATTTTGCTAATTCTTTGCCTTGCCTGTATGATTTA
961       971       981       991       1001      1011      1021      1031
TTGGATGTTAATGCTACTACTATTAGTAGAATTGATGCCACCTTTTCAGCTCGCGCCCCAAATGAAAATATAGCTAAACA
1041      1051      1061      1071      1081      1091      1101      1111
GGTTATTGACCATTTGCGAAATGTATCTAATGGTCAAACTAAATCTACTCGTTCGCAGAATTGGGAATCAACTGTTATAT
1121      1131      1141      1151      1161      1171      1181      1191
GGAATGAAACTTCCAGACACCGTACTTTAGTTGCATATTTAAAACATGTTGAGCTACAGCATTATATTCAGCAATTAAGC
1201      1211      1221      1231      1241      1251      1261      1271
TCTAAGCCATCCGCAAAAATGACCTCTTATCAAAAGGAGCAATTAAAGGTACTCTCTAATCCTGACCTGTTGGAGTTTGC
1281      1291      1301      1311      1321      1331      1341      1351
TTCCGGTCTGGTTCGCTTTGAAGCTCGAATTAAAACGCGATATTTGAAGTCTTTCGGGCTTCCTCTTAATCTTTTTGATG
1361      1371      1381      1391      1401      1411      1421      1431
CAATCCGCTTTGCTTCTGACTATAATAGTCAGGGTAAAGACCTGATTTTTGATTTATGGTCATTCTCGTTTTCTGAACTG
1441      1451      1461      1471      1481      1491      1501      1511
TTTAAAGCATTTGAGGGGGATTCAATGAATATTTATGACGATTCCGCAGTATTGGACGCTATCCAGTCTAAACATTTTAC
1521      1531      1541      1551      1561      1571      1581      1591
TATTACCCCCTCTGGCAAAACTTCTTTTGCAAAAGCCTCTCGCTATTTTGGTTTTTATCGTCGTCTGGTAAACGAGGGTT
1601      1611      1621      1631      1641      1651      1661      1671
ATGATAGTGTTGCTCTTACTATGCCTCGTAATTCCTTTTGGCGTTATGTATCTGCATTAGTTGAATGTGGTATTCCTAAA
1681      1691      1701      1711      1721      1731      1741      1751
TCTCAACTGATGAATCTTTCTACCTGTAATAATGTTGTTCCGTTAGTTCGTTTTATTAACGTAGATTTTTCTTCCCAACG
1761      1771      1781      1791      1801      1811      1821      1831
TCCTGACTGGTATAATGAGCCAGTTCTTAAAATCGCATAAGGTAATTCACAATGATTAAAGTTGAAATTAAACCATCTCA
1841      1851      1861      1871      1881      1891      1901      1911
AGCCCAATTTACTACTCGTTCTGGTGTTTCTCGTCAGGGCAAGCCTTATTCACTGAATGAGCAGCTTTGTTACGTTGATT
1921      1931      1941      1951      1961      1971      1981      1991
TGGGTAATGAATATCCGGTTCTTGTCAAGATTACTCTTGATGAAGGTCAGCCAGCCTATGCGCCTGGTCTGTACACCGTT
2001      2011      2021      2031      2041      2051      2061      2071
CATCTGTCCTCTTTCAAAGTTGGTCAGTTCGGTTCCCTTATGATTGACCGTCTGCGCCTCGTTCCGGCTAAGTAACATGG
2081      2091      2101      2111      2121      2131      2141      2151
AGCAGGTCGCGGATTTCGACACAATTTATCAGGCGATGATACAAATCTCCGTTGTACTTTGTTTCGCGCTTGGTATAATC
2161      2171      2181      2191      2201      2211      2221      2231
GCTGGGGGTCAAAGATGAGTGTTTTAGTGTATTCTTTTGCCTCTTTCGTTTTAGGTTGGTGCCTTCGTAGTGGCATTACG
2241      2251      2261      2271      2281      2291      2301      2311
TATTTTACCCGTTTAATGGAAACTTCCTCATGAAAAAGTCTTTAGTCCTCAAAGCCTCTGTAGCCGTTGCTACCCTCGTT
2321      2331      2341      2351      2361      2371      2381      2391
CCGATGCTGTCTTTCGCTGCTGAGGGTGACGATCCCGCAAAAGCGGCCTTTAACTCCCTGCAAGCCTCAGCGACCGAATA
2401      2411      2421      2431      2441      2451      2461      2471
TATCGGTTATGCGTGGGCGATGGTTGTTGTCATTGTCGGCGCAACTATCGGTATCAAGCTGTTTAAGAAATTCACCTCGA
2481      2491      2501      2511      2521      2531      2541      2551
AAGCAAGCTGATAAACCGATACAATTAAAGGCTCCTTTTGGAGCCTTTTTTTTGGAGATTTTCAACGTGAAAAAATTATT
2561      2571      2581      2591      2601      2611      2621      2631
ATTCGCAATTCCTTTAGTTGTTCCTTTCTATTCTCACTCCGCTGAAACTGTTGAAAGTTGTTTAGCAAAATCCCATACAG
2641      2651      2661      2671      2681      2691      2701      2711
AAAATTCATTTACTAACGTCTGGAAAGACGACAAAACTTTAGATCGTTACGCTAACTATGAGGGCTGTCTGTGGAATGCT
2721      2731      2741      2751      2761      2771      2781      2791
ACAGGCGTTGTAGTTTGTACTGGTGACGAAACTCAGTGTTACGGTACATGGGTTCCTATTGGGCTTGCTATCCCTGAAAA
2801      2811      2821      2831      2841      2851      2861      2871
TGAGGGTGGTGGCTCTGAGGGTGGCGGTTCTGAGGGTGGCGGTTCTGAGGGTGGCGGTACTAAACCTCCTGAGTACGGTG
2881      2891      2901      2911      2921      2931      2941      2951
ATACACCTATTCCGGGCTATACTTATATCAACCCTCTCGACGGCACTTATCCGCCTGGTACTGAGCAAAACCCCGCTAAT
2961      2971      2981      2991      3001      3011      3021      3031
CCTAATCCTTCTCTTGAGGAGTCTCAGCCTCTTAATACTTTCATGTTTCAGAATAATAGGTTCCGAAATAGGCAGGGGGC
3041      3051      3061      3071      3081      3091      3101      3111
ATTAACTGTTTATACGGGCACTGTTACTCAAGGCACTGACCCCGTTAAAACTTATTACCAGTACACTCCTGTATCATCAA
3121      3131      3141      3151      3161      3171      3181      3191
AAGCCATGTATGACGCTTACTGGAACGGTAAATTCAGAGACTGCGCTTTCCATTCTGGCTTTAATGAGGATTTATTTGTT
3201      3211      3221      3231      3241      3251      3261      3271
TGTGAATATCAAGGCCAATCGTCTGACCTGCCTCAACCTCCTGTCAATGCTGGCGGCGGCTCTGGTGGTGGTTCTGGTGG
3281      3291      3301      3311      3321      3331      3341      3351
CGGCTCTGAGGGTGGTGGCTCTGAGGGTGGCGGTTCTGAGGGTGGCGGCTCTGAGGGAGGCGGTTCCGGTGGTGGCTCTG
3361      3371      3381      3391      3401      3411      3421      3431
GTTCCGGTGATTTTGATTATGAAAAGATGGCAAACGCTAATAAGGGGGCTATGACCGAAAATGCCGATGAAAACGCGCTA
3441      3451      3461      3471      3481      3491      3501      3511
CAGTCTGACGCTAAAGGCAAACTTGATTCTGTCGCTACTGATTACGGTGCTGCTATCGATGGTTTCATTGGTGACGTTTC
3521      3531      3541      3551      3561      3571      3581      3591
CGGCCTTGCTAATGGTAATGGTGCTACTGGTGATTTTGCTGGCTCTAATTCCCAAATGGCTCAAGTCGGTGACGGTGATA
3601      3611      3621      3631      3641      3651      3661      3671
ATTCACCTTTAATGAATAATTTCCGTCAATATTTACCTTCCCTCCCTCAATCGGTTGAATGTCGCCCTTTTGTCTTTGGC
3681      3691      3701      3711      3721      3731      3741      3751
GCTGGTAAACCATATGAATTTTCTATTGATTGTGACAAAATAAACTTATTCCGTGGTGTCTTTGCGTTTCTTTTATATGT
3761      3771      3781      3791      3801      3811      3821      3831
TGCCACCTTTATGTATGTATTTTCTACGTTTGCTAACATACTGCGTAATAAGGAGTCTTAATCATGCCAGTTCTTTTGGG
3841      3851      3861      3871      3881      3891      3901      3911
TATTCCGTTATTATTGCGTTTCCTCGGTTTCCTTCTGGTAACTTTGTTCGGCTATCTGCTTACTTTTCTTAAAAAGGGCT
3921      3931      3941      3951      3961      3971      3981      3991
TCGGTAAGATAGCTATTGCTATTTCATTGTTTCTTGCTCTTATTATTGGGCTTAACTCAATTCTTGTGGGTTATCTCTCT
4001      4011      4021      4031      4041      4051      4061      4071
GATATTAGCGCTCAATTACCCTCTGACTTTGTTCAGGGTGTTCAGTTAATTCTCCCGTCTAATGCGCTTCCCTGTTTTTA
4081      4091      4101      4111      4121      4131      4141      4151
TGTTATTCTCTCTGTAAAGGCTGCTATTTTCATTTTTGACGTTAAACAAAAAATCGTTTCTTATTTGGATTGGGATAAAT
4161      4171      4181      4191      4201      4211      4221      4231
AATATGGCTGTTTATTTTGTAACTGGCAAATTAGGCTCTGGAAAGACGCTCGTTAGCGTTGGTAAGATTCAGGATAAAAT
4241      4251      4261      4271      4281      4291      4301      4311
TGTAGCTGGGTGCAAAATAGCAACTAATCTTGATTTAAGGCTTCAAAACCTCCCGCAAGTCGGGAGGTTCGCTAAAACGC
4321      4331      4341      4351      4361      4371      4381      4391
CTCGCGTTCTTAGAATACCGGATAAGCCTTCTATATCTGATTTGCTTGCTATTGGGCGCGGTAATGATTCCTACGATGAA
4401      4411      4421      4431      4441      4451      4461      4471
AATAAAAACGGCTTGCTTGTTCTCGATGAGTGCGGTACTTGGTTTAATACCCGTTCTTGGAATGATAAGGAAAGACAGCC
4481      4491      4501      4511      4521      4531      4541      4551
GATTATTGATTGGTTTCTACATGCTCGTAAATTAGGATGGGATATTATTTTTCTTGTTCAGGACTTATCTATTGTTGATA
4561      4571      4581      4591      4601      4611      4621      4631
AACAGGCGCGTTCTGCATTAGCTGAACATGTTGTTTATTGTCGTCGTCTGGACAGAATTACTTTACCTTTTGTCGGTACT
4641      4651      4661      4671      4681      4691      4701      4711
TTATATTCTCTTATTACTGGCTCGAAAATGCCTCTGCCTAAATTACATGTTGGCGTTGTTAAATATGGCGATTCTCAATT
4721      4731      4741      4751      4761      4771      4781      4791
AAGCCCTACTGTTGAGCGTTGGCTTTATACTGGTAAGAATTTGTATAACGCATATGATACTAAACAGGCTTTTTCTAGTA
4801      4811      4821      4831      4841      4851      4861      4871
ATTATGATTCCGGTGTTTATTCTTATTTAACGCCTTATTTATCACACGGTCGGTATTTCAAACCATTAAATTTAGGTCAG
4881      4891      4901      4911      4921      4931      4941      4951
AAGATGAAATTAACTAAAATATATTTGAAAAAGTTTTCTCGCGTTCTTTGTCTTGCGATTGGATTTGCATCAGCATTTAC
4961      4971      4981      4991      5001      5011      5021      5031
ATATAGTTATATAACCCAACCTAAGCCGGAGGTTAAAAAGGTAGTCTCTCAGACCTATGATTTTGATAAATTCACTATTG
5041      5051      5061      5071      5081      5091      5101      5111
ACTCTTCTCAGCGTCTTAATCTAAGCTATCGCTATGTTTTCAAGGATTCTAAGGGAAAATTAATTAATAGCGACGATTTA
5121      5131      5141      5151      5161      5171      5181      5191
CAGAAGCAAGGTTATTCACTCACATATATTGATTTATGTACTGTTTCCATTAAAAAAGGTAATTCAAATGAAATTGTTAA
5201      5211      5221      5231      5241      5251      5261      5271
ATGTAATTAATTTTGTTTTCTTGATGTTTGTTTCATCATCTTCTTTTGCTCAGGTAATTGAAATGAATAATTCGCCTCTG
5281      5291      5301      5311      5321      5331      5341      5351
CGCGATTTTGTAACTTGGTATTCAAAGCAATCAGGCGAATCCGTTATTGTTTCTCCCGATGTAAAAGGTACTGTTACTGT
5361      5371      5381      5391      5401      5411      5421      5431
ATATTCATCTGACGTTAAACCTGAAAATCTACGCAATTTCTTTATTTCTGTTTTACGTGCAAATAATTTTGATATGGTAG
5441      5451      5461      5471      5481      5491      5501      5511
GTTCTAACCCTTCCATTATTCAGAAGTATAATCCAAACAATCAGGATTATATTGATGAATTGCCATCATCTGATAATCAG
5521      5531      5541      5551      5561      5571      5581      5591
GAATATGATGATAATTCCGCTCCTTCTGGTGGTTTCTTTGTTCCGCAAAATGATAATGTTACTCAAACTTTTAAAATTAA
5601      5611      5621      5631      5641      5651      5661      5671
TAACGTTCGGGCAAAGGATTTAATACGAGTTGTCGAATTGTTTGTAAAGTCTAATACTTCTAAATCCTCAAATGTATTAT
5681      5691      5701      5711      5721      5731      5741      5751
CTATTGACGGCTCTAATCTATTAGTTGTTAGTGCTCCTAAAGATATTTTAGATAACCTTCCTCAATTCCTTTCAACTGTT
5761      5771      5781      5791      5801      5811      5821      5831
GATTTGCCAACTGACCAGATATTGATTGAGGGTTTGATATTTGAGGTTCAGCAAGGTGATGCTTTAGATTTTTCATTTGC
5841      5851      5861      5871      5881      5891      5901      5911
TGCTGGCTCTCAGCGTGGCACTGTTGCAGGCGGTGTTAATACTGACCGCCTCACCTCTGTTTTATCTTCTGCTGGTGGTT
5921      5931      5941      5951      5961      5971      5981      5991
CGTTCGGTATTTTTAATGGCGATGTTTTAGGGCTATCAGTTCGCGCATTAAAGACTAATAGCCATTCAAAAATATTGTCT
6001      6011      6021      6031      6041      6051      6061      6071
GTGCCACGTATTCTTACGCTTTCAGGTCAGAAGGGTTCTATCTCTGTTGGCCAGAATGTCCCTTTTATTACTGGTCGTGT
6081      6091      6101      6111      6121      6131      6141      6151
GACTGGTGAATCTGCCAATGTAAATAATCCATTTCAGACGATTGAGCGTCAAAATGTAGGTATTTCCATGAGCGTTTTTC
6161      6171      6181      6191      6201      6211      6221      6231
CTGTTGCAATGGCTGGCGGTAATATTGTTCTGGATATTACCAGCAAGGCCGATAGTTTGAGTTCTTCTACTCAGGCAAGT
6241      6251      6261      6271      6281      6291      6301      6311
GATGTTATTACTAATCAAAGAAGTATTGCTACAACGGTTAATTTGCGTGATGGACAGACTCTTTTACTCGGTGGCCTCAC
6321      6331      6341      6351      6361      6371      6381      6391
TGATTATAAAAACACTTCTCAGGATTCTGGCGTACCGTTCCTGTCTAAAATCCCTTTAATCGGCCTCCTGTTTAGCTCCC
6401      6411      6421      6431      6441      6451      6461      6471
GCTCTGATTCTAACGAGGAAAGCACGTTATACGTGCTCGTCAAAGCAACCATAGTACGCGCCCTGTAGCGGCGCATTAAG
6481      6491      6501      6511      6521      6531      6541      6551
CGCGGCGGGTGTGGTGGTTACGCGCAGCGTGACCGCTACACTTGCCAGCGCCCTAGCGCCCGCTCCTTTCGCTTTCTTCC
6561      6571      6581      6591      6601      6611      6621      6631
CTTCCTTTCTCGCCACGTTCGCCGGCTTTCCCCGTCAAGCTCTAAATCGGGGGCTCCCTTTAGGGTTCCGATTTAGTGCT
6641      6651      6661      6671      6681      6691      6701      6711
TTACGGCACCTCGACCCCAAAAAACTTGATTTGGGTGATGGTTCACGTAGTGGGCCATCGCCCTGATAGACGGTTTTTCG
6721      6731      6741      6751      6761      6771      6781      6791
CCCTTTGACGTTGGAGTCCACGTTCTTTAATAGTGGACTCTTGTTCCAAACTGGAACAACACTCAACCCTATCTCGGGCT
6801      6811      6821      6831      6841      6851      6861      6871
ATTCTTTTGATTTATAAGGGATTTTGCCGATTTCGGAACCACCATCAAACAGGATTTTCGCCTGCTGGGGCAAACCAGCG
6881      6891      6901      6911      6921      6931      6941      6951
TGGACCGCTTGCTGCAACTCTCTCAGGGCCAGGCGGTGAAGGGCAATCAGCTGTTGCCCGTCTCACTGGTGAAAAGAAAA
6961      6971      6981      6991      7001      7011      7021      7031
ACCACCCTGGCGCCCAATACGCAAACCGCCTCTCCCCGCGCGTTGGCCGATTCATTAATGCAGCTGGCACGACAGGTTTC
7041      7051      7061      7071      7081      7091      7101      7111
CCGACTGGAAAGCGGGCAGTGAGCGCAACGCAATTAATGTGAGTTAGCTCACTCATTAGGCACCCCAGGCTTTACACTTT
7121      7131      7141      7151      7161      7171      7181      7191
ATGCTTCCGGCTCGTATGTTGTGTGGAATTGTGAGCGGATAACAATTTCACACAGGAAACAGCTATGACCATGATTACGA
7201      7211      7221      7231      7241      7251      7261      7271
ATTCGAGCTCGGTACCCGGGGATCCTCAACTGTGAGGAGGCTCACGGACGCGAAGAACAGGCACGCGTGCTGGCAGAAAC
7281      7291      7301      7311      7321      7331      7341      7351
CCCCGGTATGACCGTGAAAACGGCCCGCCGCATTCTGGCCGCAGCACCACAGAGTGCACAGGCGCGCAGTGACACTGCGC
7361      7371      7381      7391      7401      7411      7421      7431
TGGATCGTCTGATGCAGGGGGCACCGGCACCGCTGGCTGCAGGTAACCCGGCATCTGATGCCGTTAACGATTTGCTGAAC
7441      7451      7461      7471      7481      7491      7501      7511
ACACCAGTGTAAGGGATGTTTATGACGAGCAAAGAAACCTTTACCCATTACCAGCCGCAGGGCAACAGTGACCCGGCTCA
7521      7531      7541      7551      7561      7571      7581      7591
TACCGCAACCGCGCCCGGCGGATTGAGTGCGAAAGCGCCTGCAATGACCCCGCTGATGCTGGACACCTCCAGCCGTAAGC
7601      7611      7621      7631      7641      7651      7661      7671
TGGTTGCGTGGGATGGCACCACCGACGGTGCTGCCGTTGGCATTCTTGCGGTTGCTGCTGACCAGACCAGCACCACGCTG
7681      7691      7701      7711      7721      7731      7741      7751
ACGTTCTACAAGTCCGGCACGTTCCGTTATGAGGATGTGCTCTGGCCGGAGGCTGCCAGCGACGAGACGAAAAAACGGAC
7761      7771      7781      7791      7801      7811      7821      7831
CGCGTTTGCCGGAACGGCAATCAGCATCGTTTAACTTTACCCTTCATCACTAAAGGCCGCCTGTGCGGCTTTTTTTACGG
7841      7851      7861      7871      7881      7891      7901      7911
GATTTTTTTATGTCGATGTACACAACCGCCCAACTGCTGGCGGCAAATGAGCAGAAATTTAAGTTTGATCCGCTGTTTCT
7921      7931      7941      7951      7961      7971      7981      7991
GCGTCTCTTTTTCCGTGAGAGCTATCCCTTCACCACGGAGAAAGTCTATCTCTCACAAATTCCGGGACTGGTAAACATGG
8001      8011      8021      8031      8041      8051      8061
CGCTGTACGTTTCGCCGATTGTTTCCGGTGAGGTTATCCGTTCCCGTGGCGGCTCCACCTCTGA
\end{verbatim}

\subsubsection*{7,249 scaffold}
\si{7,249}~basepair scaffold version used for the extension lever structure:
\begin{verbatim}
1         11        21        31        41        51        61        71
AAGCTTGGCACTGGCCGTCGTTTTACAACGTCGTGACTGGGAAAACCCTGGCGTTACCCAACTTAATCGCCTTGCAGCAC
81        91        101       111       121       131       141       151
ATCCCCCTTTCGCCAGCTGGCGTAATAGCGAAGAGGCCCGCACCGATCGCCCTTCCCAACAGTTGCGCAGCCTGAATGGC
161       171       181       191       201       211       221       231
GAATGGCGCTTTGCCTGGTTTCCGGCACCAGAAGCGGTGCCGGAAAGCTGGCTGGAGTGCGATCTTCCTGAGGCCGATAC
241       251       261       271       281       291       301       311
TGTCGTCGTCCCCTCAAACTGGCAGATGCACGGTTACGATGCGCCCATCTACACCAACGTGACCTATCCCATTACGGTCA
321       331       341       351       361       371       381       391
ATCCGCCGTTTGTTCCCACGGAGAATCCGACGGGTTGTTACTCGCTCACATTTAATGTTGATGAAAGCTGGCTACAGGAA
401       411       421       431       441       451       461       471
GGCCAGACGCGAATTATTTTTGATGGCGTTCCTATTGGTTAAAAAATGAGCTGATTTAACAAAAATTTAATGCGAATTTT
481       491       501       511       521       531       541       551
AACAAAATATTAACGTTTACAATTTAAATATTTGCTTATACAATCTTCCTGTTTTTGGGGCTTTTCTGATTATCAACCGG
561       571       581       591       601       611       621       631
GGTACATATGATTGACATGCTAGTTTTACGATTACCGTTCATCGATTCTCTTGTTTGCTCCAGACTCTCAGGCAATGACC
641       651       661       671       681       691       701       711
TGATAGCCTTTGTAGATCTCTCAAAAATAGCTACCCTCTCCGGCATTAATTTATCAGCTAGAACGGTTGAATATCATATT
721       731       741       751       761       771       781       791
GATGGTGATTTGACTGTCTCCGGCCTTTCTCACCCTTTTGAATCTTTACCTACACATTACTCAGGCATTGCATTTAAAAT
801       811       821       831       841       851       861       871
ATATGAGGGTTCTAAAAATTTTTATCCTTGCGTTGAAATAAAGGCTTCTCCCGCAAAAGTATTACAGGGTCATAATGTTT
881       891       901       911       921       931       941       951
TTGGTACAACCGATTTAGCTTTATGCTCTGAGGCTTTATTGCTTAATTTTGCTAATTCTTTGCCTTGCCTGTATGATTTA
961       971       981       991       1001      1011      1021      1031
TTGGATGTTAATGCTACTACTATTAGTAGAATTGATGCCACCTTTTCAGCTCGCGCCCCAAATGAAAATATAGCTAAACA
1041      1051      1061      1071      1081      1091      1101      1111
GGTTATTGACCATTTGCGAAATGTATCTAATGGTCAAACTAAATCTACTCGTTCGCAGAATTGGGAATCAACTGTTATAT
1121      1131      1141      1151      1161      1171      1181      1191
GGAATGAAACTTCCAGACACCGTACTTTAGTTGCATATTTAAAACATGTTGAGCTACAGCATTATATTCAGCAATTAAGC
1201      1211      1221      1231      1241      1251      1261      1271
TCTAAGCCATCCGCAAAAATGACCTCTTATCAAAAGGAGCAATTAAAGGTACTCTCTAATCCTGACCTGTTGGAGTTTGC
1281      1291      1301      1311      1321      1331      1341      1351
TTCCGGTCTGGTTCGCTTTGAAGCTCGAATTAAAACGCGATATTTGAAGTCTTTCGGGCTTCCTCTTAATCTTTTTGATG
1361      1371      1381      1391      1401      1411      1421      1431
CAATCCGCTTTGCTTCTGACTATAATAGTCAGGGTAAAGACCTGATTTTTGATTTATGGTCATTCTCGTTTTCTGAACTG
1441      1451      1461      1471      1481      1491      1501      1511
TTTAAAGCATTTGAGGGGGATTCAATGAATATTTATGACGATTCCGCAGTATTGGACGCTATCCAGTCTAAACATTTTAC
1521      1531      1541      1551      1561      1571      1581      1591
TATTACCCCCTCTGGCAAAACTTCTTTTGCAAAAGCCTCTCGCTATTTTGGTTTTTATCGTCGTCTGGTAAACGAGGGTT
1601      1611      1621      1631      1641      1651      1661      1671
ATGATAGTGTTGCTCTTACTATGCCTCGTAATTCCTTTTGGCGTTATGTATCTGCATTAGTTGAATGTGGTATTCCTAAA
1681      1691      1701      1711      1721      1731      1741      1751
TCTCAACTGATGAATCTTTCTACCTGTAATAATGTTGTTCCGTTAGTTCGTTTTATTAACGTAGATTTTTCTTCCCAACG
1761      1771      1781      1791      1801      1811      1821      1831
TCCTGACTGGTATAATGAGCCAGTTCTTAAAATCGCATAAGGTAATTCACAATGATTAAAGTTGAAATTAAACCATCTCA
1841      1851      1861      1871      1881      1891      1901      1911
AGCCCAATTTACTACTCGTTCTGGTGTTTCTCGTCAGGGCAAGCCTTATTCACTGAATGAGCAGCTTTGTTACGTTGATT
1921      1931      1941      1951      1961      1971      1981      1991
TGGGTAATGAATATCCGGTTCTTGTCAAGATTACTCTTGATGAAGGTCAGCCAGCCTATGCGCCTGGTCTGTACACCGTT
2001      2011      2021      2031      2041      2051      2061      2071
CATCTGTCCTCTTTCAAAGTTGGTCAGTTCGGTTCCCTTATGATTGACCGTCTGCGCCTCGTTCCGGCTAAGTAACATGG
2081      2091      2101      2111      2121      2131      2141      2151
AGCAGGTCGCGGATTTCGACACAATTTATCAGGCGATGATACAAATCTCCGTTGTACTTTGTTTCGCGCTTGGTATAATC
2161      2171      2181      2191      2201      2211      2221      2231
GCTGGGGGTCAAAGATGAGTGTTTTAGTGTATTCTTTTGCCTCTTTCGTTTTAGGTTGGTGCCTTCGTAGTGGCATTACG
2241      2251      2261      2271      2281      2291      2301      2311
TATTTTACCCGTTTAATGGAAACTTCCTCATGAAAAAGTCTTTAGTCCTCAAAGCCTCTGTAGCCGTTGCTACCCTCGTT
2321      2331      2341      2351      2361      2371      2381      2391
CCGATGCTGTCTTTCGCTGCTGAGGGTGACGATCCCGCAAAAGCGGCCTTTAACTCCCTGCAAGCCTCAGCGACCGAATA
2401      2411      2421      2431      2441      2451      2461      2471
TATCGGTTATGCGTGGGCGATGGTTGTTGTCATTGTCGGCGCAACTATCGGTATCAAGCTGTTTAAGAAATTCACCTCGA
2481      2491      2501      2511      2521      2531      2541      2551
AAGCAAGCTGATAAACCGATACAATTAAAGGCTCCTTTTGGAGCCTTTTTTTTGGAGATTTTCAACGTGAAAAAATTATT
2561      2571      2581      2591      2601      2611      2621      2631
ATTCGCAATTCCTTTAGTTGTTCCTTTCTATTCTCACTCCGCTGAAACTGTTGAAAGTTGTTTAGCAAAATCCCATACAG
2641      2651      2661      2671      2681      2691      2701      2711
AAAATTCATTTACTAACGTCTGGAAAGACGACAAAACTTTAGATCGTTACGCTAACTATGAGGGCTGTCTGTGGAATGCT
2721      2731      2741      2751      2761      2771      2781      2791
ACAGGCGTTGTAGTTTGTACTGGTGACGAAACTCAGTGTTACGGTACATGGGTTCCTATTGGGCTTGCTATCCCTGAAAA
2801      2811      2821      2831      2841      2851      2861      2871
TGAGGGTGGTGGCTCTGAGGGTGGCGGTTCTGAGGGTGGCGGTTCTGAGGGTGGCGGTACTAAACCTCCTGAGTACGGTG
2881      2891      2901      2911      2921      2931      2941      2951
ATACACCTATTCCGGGCTATACTTATATCAACCCTCTCGACGGCACTTATCCGCCTGGTACTGAGCAAAACCCCGCTAAT
2961      2971      2981      2991      3001      3011      3021      3031
CCTAATCCTTCTCTTGAGGAGTCTCAGCCTCTTAATACTTTCATGTTTCAGAATAATAGGTTCCGAAATAGGCAGGGGGC
3041      3051      3061      3071      3081      3091      3101      3111
ATTAACTGTTTATACGGGCACTGTTACTCAAGGCACTGACCCCGTTAAAACTTATTACCAGTACACTCCTGTATCATCAA
3121      3131      3141      3151      3161      3171      3181      3191
AAGCCATGTATGACGCTTACTGGAACGGTAAATTCAGAGACTGCGCTTTCCATTCTGGCTTTAATGAGGATTTATTTGTT
3201      3211      3221      3231      3241      3251      3261      3271
TGTGAATATCAAGGCCAATCGTCTGACCTGCCTCAACCTCCTGTCAATGCTGGCGGCGGCTCTGGTGGTGGTTCTGGTGG
3281      3291      3301      3311      3321      3331      3341      3351
CGGCTCTGAGGGTGGTGGCTCTGAGGGTGGCGGTTCTGAGGGTGGCGGCTCTGAGGGAGGCGGTTCCGGTGGTGGCTCTG
3361      3371      3381      3391      3401      3411      3421      3431
GTTCCGGTGATTTTGATTATGAAAAGATGGCAAACGCTAATAAGGGGGCTATGACCGAAAATGCCGATGAAAACGCGCTA
3441      3451      3461      3471      3481      3491      3501      3511
CAGTCTGACGCTAAAGGCAAACTTGATTCTGTCGCTACTGATTACGGTGCTGCTATCGATGGTTTCATTGGTGACGTTTC
3521      3531      3541      3551      3561      3571      3581      3591
CGGCCTTGCTAATGGTAATGGTGCTACTGGTGATTTTGCTGGCTCTAATTCCCAAATGGCTCAAGTCGGTGACGGTGATA
3601      3611      3621      3631      3641      3651      3661      3671
ATTCACCTTTAATGAATAATTTCCGTCAATATTTACCTTCCCTCCCTCAATCGGTTGAATGTCGCCCTTTTGTCTTTGGC
3681      3691      3701      3711      3721      3731      3741      3751
GCTGGTAAACCATATGAATTTTCTATTGATTGTGACAAAATAAACTTATTCCGTGGTGTCTTTGCGTTTCTTTTATATGT
3761      3771      3781      3791      3801      3811      3821      3831
TGCCACCTTTATGTATGTATTTTCTACGTTTGCTAACATACTGCGTAATAAGGAGTCTTAATCATGCCAGTTCTTTTGGG
3841      3851      3861      3871      3881      3891      3901      3911
TATTCCGTTATTATTGCGTTTCCTCGGTTTCCTTCTGGTAACTTTGTTCGGCTATCTGCTTACTTTTCTTAAAAAGGGCT
3921      3931      3941      3951      3961      3971      3981      3991
TCGGTAAGATAGCTATTGCTATTTCATTGTTTCTTGCTCTTATTATTGGGCTTAACTCAATTCTTGTGGGTTATCTCTCT
4001      4011      4021      4031      4041      4051      4061      4071
GATATTAGCGCTCAATTACCCTCTGACTTTGTTCAGGGTGTTCAGTTAATTCTCCCGTCTAATGCGCTTCCCTGTTTTTA
4081      4091      4101      4111      4121      4131      4141      4151
TGTTATTCTCTCTGTAAAGGCTGCTATTTTCATTTTTGACGTTAAACAAAAAATCGTTTCTTATTTGGATTGGGATAAAT
4161      4171      4181      4191      4201      4211      4221      4231
AATATGGCTGTTTATTTTGTAACTGGCAAATTAGGCTCTGGAAAGACGCTCGTTAGCGTTGGTAAGATTCAGGATAAAAT
4241      4251      4261      4271      4281      4291      4301      4311
TGTAGCTGGGTGCAAAATAGCAACTAATCTTGATTTAAGGCTTCAAAACCTCCCGCAAGTCGGGAGGTTCGCTAAAACGC
4321      4331      4341      4351      4361      4371      4381      4391
CTCGCGTTCTTAGAATACCGGATAAGCCTTCTATATCTGATTTGCTTGCTATTGGGCGCGGTAATGATTCCTACGATGAA
4401      4411      4421      4431      4441      4451      4461      4471
AATAAAAACGGCTTGCTTGTTCTCGATGAGTGCGGTACTTGGTTTAATACCCGTTCTTGGAATGATAAGGAAAGACAGCC
4481      4491      4501      4511      4521      4531      4541      4551
GATTATTGATTGGTTTCTACATGCTCGTAAATTAGGATGGGATATTATTTTTCTTGTTCAGGACTTATCTATTGTTGATA
4561      4571      4581      4591      4601      4611      4621      4631
AACAGGCGCGTTCTGCATTAGCTGAACATGTTGTTTATTGTCGTCGTCTGGACAGAATTACTTTACCTTTTGTCGGTACT
4641      4651      4661      4671      4681      4691      4701      4711
TTATATTCTCTTATTACTGGCTCGAAAATGCCTCTGCCTAAATTACATGTTGGCGTTGTTAAATATGGCGATTCTCAATT
4721      4731      4741      4751      4761      4771      4781      4791
AAGCCCTACTGTTGAGCGTTGGCTTTATACTGGTAAGAATTTGTATAACGCATATGATACTAAACAGGCTTTTTCTAGTA
4801      4811      4821      4831      4841      4851      4861      4871
ATTATGATTCCGGTGTTTATTCTTATTTAACGCCTTATTTATCACACGGTCGGTATTTCAAACCATTAAATTTAGGTCAG
4881      4891      4901      4911      4921      4931      4941      4951
AAGATGAAATTAACTAAAATATATTTGAAAAAGTTTTCTCGCGTTCTTTGTCTTGCGATTGGATTTGCATCAGCATTTAC
4961      4971      4981      4991      5001      5011      5021      5031
ATATAGTTATATAACCCAACCTAAGCCGGAGGTTAAAAAGGTAGTCTCTCAGACCTATGATTTTGATAAATTCACTATTG
5041      5051      5061      5071      5081      5091      5101      5111
ACTCTTCTCAGCGTCTTAATCTAAGCTATCGCTATGTTTTCAAGGATTCTAAGGGAAAATTAATTAATAGCGACGATTTA
5121      5131      5141      5151      5161      5171      5181      5191
CAGAAGCAAGGTTATTCACTCACATATATTGATTTATGTACTGTTTCCATTAAAAAAGGTAATTCAAATGAAATTGTTAA
5201      5211      5221      5231      5241      5251      5261      5271
ATGTAATTAATTTTGTTTTCTTGATGTTTGTTTCATCATCTTCTTTTGCTCAGGTAATTGAAATGAATAATTCGCCTCTG
5281      5291      5301      5311      5321      5331      5341      5351
CGCGATTTTGTAACTTGGTATTCAAAGCAATCAGGCGAATCCGTTATTGTTTCTCCCGATGTAAAAGGTACTGTTACTGT
5361      5371      5381      5391      5401      5411      5421      5431
ATATTCATCTGACGTTAAACCTGAAAATCTACGCAATTTCTTTATTTCTGTTTTACGTGCAAATAATTTTGATATGGTAG
5441      5451      5461      5471      5481      5491      5501      5511
GTTCTAACCCTTCCATTATTCAGAAGTATAATCCAAACAATCAGGATTATATTGATGAATTGCCATCATCTGATAATCAG
5521      5531      5541      5551      5561      5571      5581      5591
GAATATGATGATAATTCCGCTCCTTCTGGTGGTTTCTTTGTTCCGCAAAATGATAATGTTACTCAAACTTTTAAAATTAA
5601      5611      5621      5631      5641      5651      5661      5671
TAACGTTCGGGCAAAGGATTTAATACGAGTTGTCGAATTGTTTGTAAAGTCTAATACTTCTAAATCCTCAAATGTATTAT
5681      5691      5701      5711      5721      5731      5741      5751
CTATTGACGGCTCTAATCTATTAGTTGTTAGTGCTCCTAAAGATATTTTAGATAACCTTCCTCAATTCCTTTCAACTGTT
5761      5771      5781      5791      5801      5811      5821      5831
GATTTGCCAACTGACCAGATATTGATTGAGGGTTTGATATTTGAGGTTCAGCAAGGTGATGCTTTAGATTTTTCATTTGC
5841      5851      5861      5871      5881      5891      5901      5911
TGCTGGCTCTCAGCGTGGCACTGTTGCAGGCGGTGTTAATACTGACCGCCTCACCTCTGTTTTATCTTCTGCTGGTGGTT
5921      5931      5941      5951      5961      5971      5981      5991
CGTTCGGTATTTTTAATGGCGATGTTTTAGGGCTATCAGTTCGCGCATTAAAGACTAATAGCCATTCAAAAATATTGTCT
6001      6011      6021      6031      6041      6051      6061      6071
GTGCCACGTATTCTTACGCTTTCAGGTCAGAAGGGTTCTATCTCTGTTGGCCAGAATGTCCCTTTTATTACTGGTCGTGT
6081      6091      6101      6111      6121      6131      6141      6151
GACTGGTGAATCTGCCAATGTAAATAATCCATTTCAGACGATTGAGCGTCAAAATGTAGGTATTTCCATGAGCGTTTTTC
6161      6171      6181      6191      6201      6211      6221      6231
CTGTTGCAATGGCTGGCGGTAATATTGTTCTGGATATTACCAGCAAGGCCGATAGTTTGAGTTCTTCTACTCAGGCAAGT
6241      6251      6261      6271      6281      6291      6301      6311
GATGTTATTACTAATCAAAGAAGTATTGCTACAACGGTTAATTTGCGTGATGGACAGACTCTTTTACTCGGTGGCCTCAC
6321      6331      6341      6351      6361      6371      6381      6391
TGATTATAAAAACACTTCTCAGGATTCTGGCGTACCGTTCCTGTCTAAAATCCCTTTAATCGGCCTCCTGTTTAGCTCCC
6401      6411      6421      6431      6441      6451      6461      6471
GCTCTGATTCTAACGAGGAAAGCACGTTATACGTGCTCGTCAAAGCAACCATAGTACGCGCCCTGTAGCGGCGCATTAAG
6481      6491      6501      6511      6521      6531      6541      6551
CGCGGCGGGTGTGGTGGTTACGCGCAGCGTGACCGCTACACTTGCCAGCGCCCTAGCGCCCGCTCCTTTCGCTTTCTTCC
6561      6571      6581      6591      6601      6611      6621      6631
CTTCCTTTCTCGCCACGTTCGCCGGCTTTCCCCGTCAAGCTCTAAATCGGGGGCTCCCTTTAGGGTTCCGATTTAGTGCT
6641      6651      6661      6671      6681      6691      6701      6711
TTACGGCACCTCGACCCCAAAAAACTTGATTTGGGTGATGGTTCACGTAGTGGGCCATCGCCCTGATAGACGGTTTTTCG
6721      6731      6741      6751      6761      6771      6781      6791
CCCTTTGACGTTGGAGTCCACGTTCTTTAATAGTGGACTCTTGTTCCAAACTGGAACAACACTCAACCCTATCTCGGGCT
6801      6811      6821      6831      6841      6851      6861      6871
ATTCTTTTGATTTATAAGGGATTTTGCCGATTTCGGAACCACCATCAAACAGGATTTTCGCCTGCTGGGGCAAACCAGCG
6881      6891      6901      6911      6921      6931      6941      6951
TGGACCGCTTGCTGCAACTCTCTCAGGGCCAGGCGGTGAAGGGCAATCAGCTGTTGCCCGTCTCACTGGTGAAAAGAAAA
6961      6971      6981      6991      7001      7011      7021      7031
ACCACCCTGGCGCCCAATACGCAAACCGCCTCTCCCCGCGCGTTGGCCGATTCATTAATGCAGCTGGCACGACAGGTTTC
7041      7051      7061      7071      7081      7091      7101      7111
CCGACTGGAAAGCGGGCAGTGAGCGCAACGCAATTAATGTGAGTTAGCTCACTCATTAGGCACCCCAGGCTTTACACTTT
7121      7131      7141      7151      7161      7171      7181      7191
ATGCTTCCGGCTCGTATGTTGTGTGGAATTGTGAGCGGATAACAATTTCACACAGGAAACAGCTATGACCATGATTACGA
7201      7211      7221      7231      7241
ATTCGAGCTCGGTACCCGGGGATCCTCTAGAGTCGACCTGCAGGCATGC
\end{verbatim}

\subsection*{Staple sequences}
All used staple sequences for the assembly of a bistable switch structure with extension lever. Names follow the nomenclature 5'-helix[base]3'-helix[base].

\begin{longtable}{lll}
\caption{Bistable switch stator sequences} \\
\toprule
Type & Name & Sequence \\
\midrule
\endfirsthead

\toprule
Type & Name & Sequence \\
\midrule
\endhead

\midrule
\multicolumn{3}{r}{{Continued on next page}} \\
\midrule
\endfoot

\bottomrule
\endlastfoot
Basic\_stator          & 1{[}56{]}2{[}72{]}     & GTCAGGACGTTGGGAAGAAAAATCTAATAGTA                  \\
Basic\_stator          & 1{[}88{]}2{[}104{]}    & ATAAAACGAACTAACGGAACAACAGCTTTTGC                  \\
Basic\_stator          & 1{[}152{]}2{[}168{]}   & ATACCACATTCAACTAATGCAGATGTAAGAGC                  \\
Basic\_stator          & 1{[}120{]}2{[}136{]}   & GGTAGAAAGATTCATCAGTTGAGCAGACGAC                   \\
Basic\_stator          & 2{[}167{]}26{[}40{]}   & AACACTATATCGCGTTCAAAGAACTAATTTCATCTTCTGAACAACATG  \\
Basic\_stator          & 2{[}39{]}42{[}40{]}    & GGAATCGTTTAAACTCATCTTTAAAGTACAACGGAGATCTTGAGAT    \\
Basic\_stator          & 2{[}135{]}30{[}40{]}   & GATAAAAAGATTAACAATTTTATTTCCAGAGCCTAATTAAAGTAAG    \\
Basic\_stator          & 2{[}71{]}38{[}40{]}    & AAATGTTTAAATCAAAGGTTTAGTACCGCCACCCTCAGAGAAAAAGGC  \\
\end{longtable}

\begin{longtable}{lll}
\caption{Bistable switch rotor sequences} \\
\toprule
Type & Name & Sequence \\
\midrule
\endfirsthead

\toprule
Type & Name & Sequence \\
\midrule
\endhead

\midrule
\multicolumn{3}{r}{{Continued on next page}} \\
\midrule
\endfoot

\bottomrule
\endlastfoot
1st rotor                 & 1[193]3[201] & AGCGGTGCCGCATACCGAGTTGA                                       \\
1st rotor                 & 5[196]4[189] & AATGGGTGCGCGGTTGCGGTA                                         \\
1st rotor                 & 3[202]3[225] & GGATCCCCGGGTACCGACAACCAG                                      \\
1st rotor                 & 1[217]2[209] & AGACGATGCGGCGGGCCGTTTT                                        \\
1st rotor                 & 2[208]0[196] & CACGGTGTGCCCCCGTTAACGGCATCA                                   \\
1st rotor                 & 4[216]1[216] & ACTCAATCCGCCGGAAAGGTTTCTTTGCAGCAAATCTGCATC                    \\
Connector extension & 2[261]3[246] & TTAAAGGCCGCTGCACTCTGTGGTGGTGTCCAG                             \\
Connector extension & 2[237]1[261] & GCTGCGGCCAGAATCCAGCGCAGTGTCACTGCGCGCCTGTTTTGCGGGATCG          \\
\end{longtable}

\begin{longtable}{lll}
\caption{Extension sequences} \\
\toprule
Type & Name & Sequence \\
\midrule
\endfirsthead

\toprule
Type & Name & Sequence \\
\midrule
\endhead

\midrule
\multicolumn{3}{r}{{Continued on next page}} \\
\midrule
\endfoot

\bottomrule
\endlastfoot
Basic extension  & 1{[}154{]}3{[}167{]}       & GCAACGGCGCCCACGCCTTGAGTAACAG                                                   \\
Basic extension  & 0{[}832{]}3{[}825{]}       & AAGCAAACGAGCCGAAACGCAATAATAAAGAATAA                                            \\
Basic extension  & 1{[}553{]}3{[}566{]}       & AACAAAGAACAAACGAATCCTTGAAAAC                                                   \\
Basic extension  & 1{[}1036{]}3{[}1049{]}     & CTTTGATAGCACTACGACGACAATAAAC                                                   \\
Basic extension  & 0{[}643{]}3{[}636{]}       & TATAATGCTGCAAGCAAAAGGGCGACATCTGAGAG                                            \\
Basic extension  & 1{[}280{]}3{[}293{]}       & AAACACTTTGAAAATTGCTCAGTACCAG                                                   \\
Basic extension  & 1{[}343{]}3{[}356{]}       & TATCATCGTTTCAGTGTATCACCGTACT                                                   \\
Basic extension  & 1{[}1015{]}3{[}1028{]}     & TTAACCGAGAGCCGGGTAAAGTAATTCT                                                   \\
\end{longtable}

\begin{longtable}{lll}
\caption{Extension NO AuNR staples and AuNR binder staples} \\
\toprule
Type & Name & Sequence \\
\midrule
\endfirsthead

\toprule
Type & Name & Sequence \\
\midrule
\endhead

\midrule
\multicolumn{3}{r}{{Continued on next page}} \\
\midrule
\endfoot

\bottomrule
\endlastfoot
Basic\_no\_AuNR  & 2{[}195{]}0{[}182{]}       & ATACCGATAGTTGCAAGACTTAAATCAC                                                   \\
Basic\_no\_AuNR  & 4{[}188{]}1{[}195{]}       & TGGCCTTACAGGAAGATTGTAGACAGTCTTTCATG                                            \\
Basic\_no\_AuNR  & 4{[}230{]}1{[}237{]}       & CCGCCGCATATTTTGTTAAAAGTAATGTCCACTAC                                            \\
Basic\_no\_AuNR  & 2{[}237{]}0{[}224{]}       & GTTTATCAGCTTGCCGTAATGGTAGGTA                                                   \\
Basic\_no\_AuNR  & 2{[}321{]}0{[}308{]}       & CAACTAAAGGAATTATACCAATTTGCGG                                                   \\
Basic\_no\_AuNR  & 4{[}251{]}1{[}258{]}       & CCACCAGTAAATTTTTGTTAATATTTTAAAAACGA                                            \\
Basic\_no\_AuNR  & 2{[}279{]}0{[}266{]}       & TCTCCAAAAAAAAGAAAAGAATTAGAAC                                                   \\
Basic\_no\_AuNR  & 2{[}300{]}0{[}287{]}       & ATAATTTTTTCACGCATCTTTTCAACGC                                                   \\
\end{longtable}

\clearpage
\section{oxDNA switching simulations}

To gain deeper insight into the possible switching mechanism of our bistable switch structure, we performed oxDNA simulations under applied external forces. 

\begin{figure}[htbp!]
\centering
\includegraphics[width=0.35\textwidth]{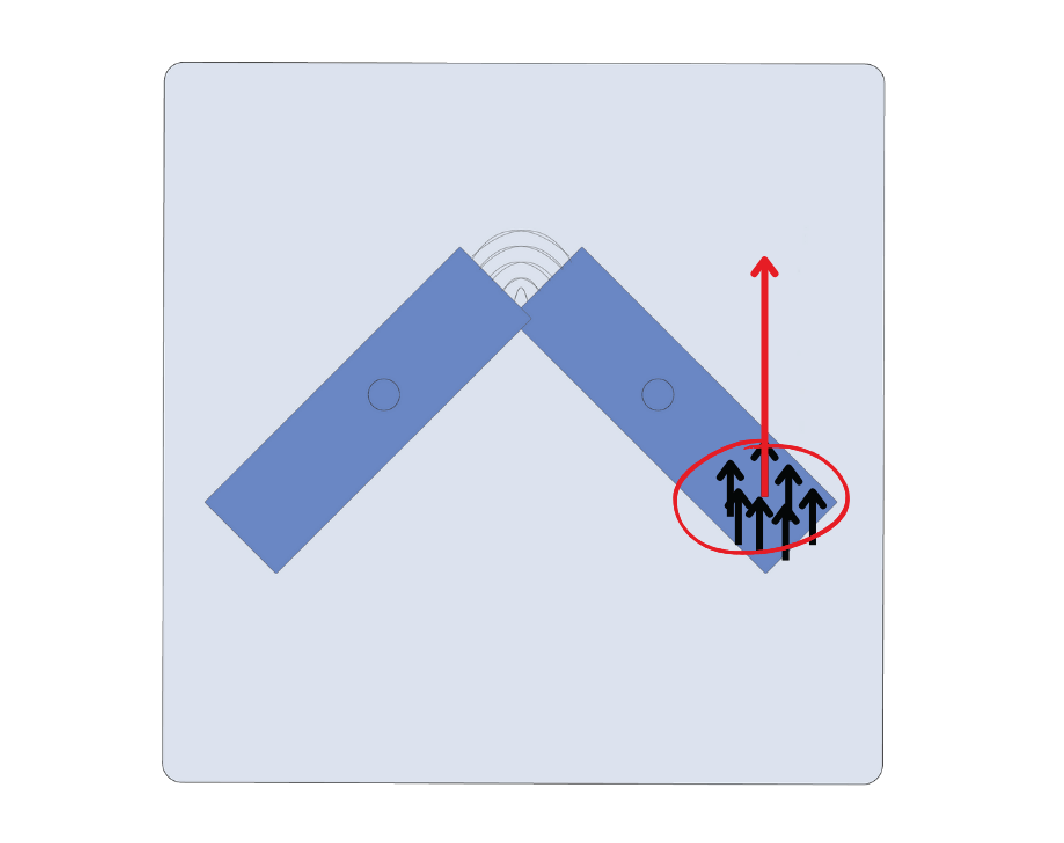}
\caption{Visualization of the applied force on the rotor arm. The effective force was distributed among \si{230} nucleotides on the right rotor arm.}
\label{SI12}
\end{figure}

We simulated the structure in response to a constant effective force of \SI{39.1}{\pico\newton} on the right rotor arm (cf. Supplemental Figure~\ref{SI12}). The magnitude corresponds well to electric fields with strengths up to \SI{732}{\volt\per\centi\meter} (\SI{300}{\volt}) applied to the lever arm extension during experiments, which generates a maximum torque of $\SI{188}{\pico\newton\nano\meter}$ \cite{vogt2023electrokinetic}. The effective force was distributed among \si{230} nucleotides at the lower‐right end of the right 6HB rotor arm (see Supplemental Figure~\ref{SI12}, arrows), yielding approximately \SI{0.17}{\pico\newton} per nucleotide.
\\

\begin{figure}[htbp!]
\centering
\includegraphics[width=1\textwidth]{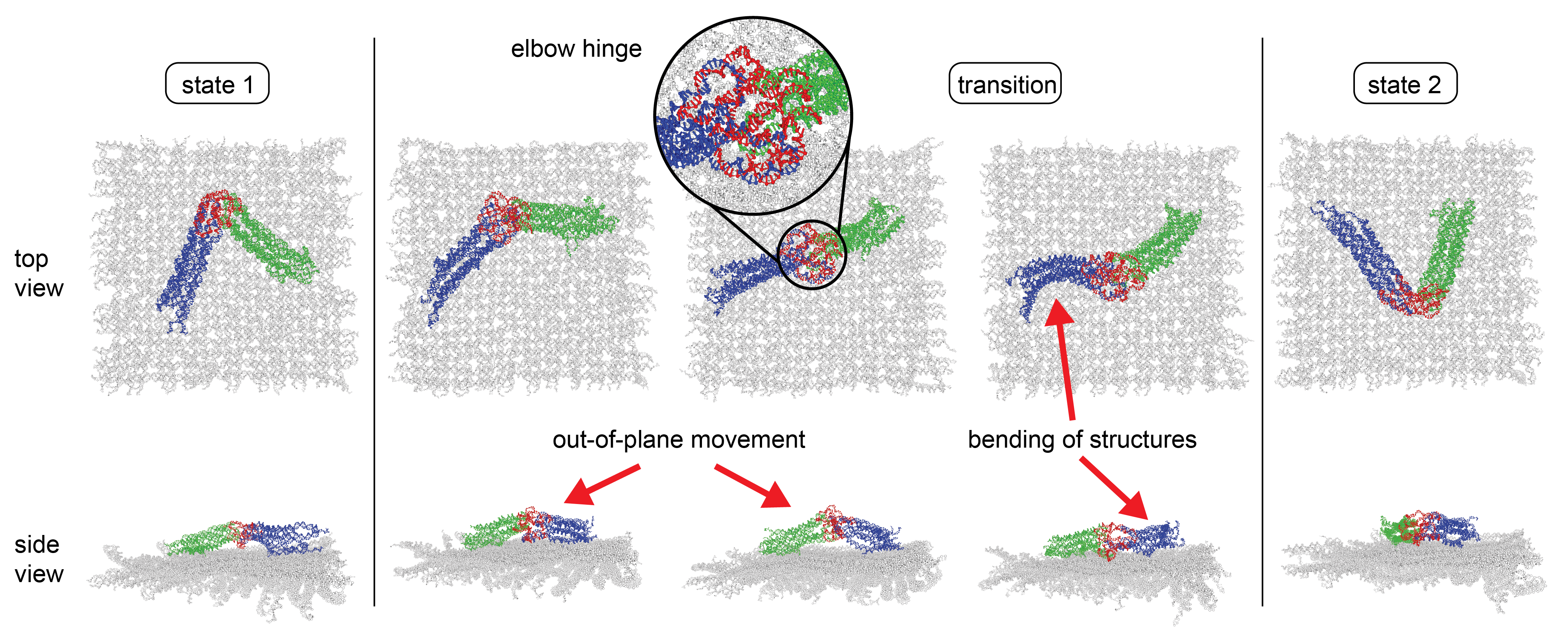}
\caption{Snapshots of the oxDNA simulation with applied external force equivalent to \SI{39.1}{\pico\newton} external field at the right rotor. Snapshots of the top and side view show typical characteristics during transition. Bending and out-of-plane movement are apparent features.}
\label{SI13}
\end{figure}

The simulations reveal that the 6HB rotor arms show both bending as well as a compensatory out-of-plane movement during transition (cf. Supplemental Figure~\ref{SI13} red arrows). Due to the low number of crossovers in the elbow hinge region, the local bending stiffness is reduced, causing the rotors to behave more like three independent two-helix bundles and enabling them to pass (through) one another (cf. Supplemental Figure~\ref{SI13} zoom lens elbow hinge).

\clearpage
\section{Theoretical considerations} 
\label{sec:theory}
In the following, we present a semi-quantitative analysis of the bistable switching behavior observed in our system. Several model parameters are only approximately known—most notably, the coupling factor $\xi$, which accounts for the conversion of applied voltage into mechanical energy or torque, and the effective retardation coefficient $\gamma_R$, which characterizes the Brownian movement of the switch. Additionally, the precise shape of the underlying energy landscape remains unknown. As the model is constructed on general and physically reasonable assumptions and reproduces the key qualitative features of the system's behavior, it is nevertheless assumed to provide good insight into the underlying physical mechanisms at play.

\subsection*{Quartic potential}
We model the mechanical potential energy landscape of the  bistable switch with the \emph{quartic double-well potential}:
\begin{equation}
E(\varphi)=A \varphi^4 - B\varphi^2 + C \quad
  A>0,\quad B>0,\quad C\in\mathbb{R}.
  \label{eq:E_varphi}
\end{equation}

This potential has two local minima at nonzero $\varphi$ and one local maximum at $\varphi=0$.  

From
\begin{equation}
  \frac{dE}{d\varphi} =  4 A \varphi^3 - 2 B \varphi = 0
\end{equation}

we get 
\[
  \varphi \;=\; 0
  \quad\text{or}
  \quad \varphi_{\pm} 
  \;=\;
  \pm\,\sqrt{\frac{B}{2A}}.
\]

At the minima the energy is: 
\[
 E(\varphi_{\pm}) = A \varphi_{\pm}^4 - B \varphi_{\pm}^2 + C = A \frac{B^2}{4A^2} - B \frac{B}{2A} + C = C - \frac{B^2}{4A}, 
\]
whereas at the maximum we have  

\[
E(0) = C.
\]

This results in a barrier height of 
\[
  \Delta E =   E(0) - E(\varphi_\pm) = \frac{B^2}{4 A}
\]

\begin{figure}[htb]
\centering
\includegraphics[width=0.5\textwidth]{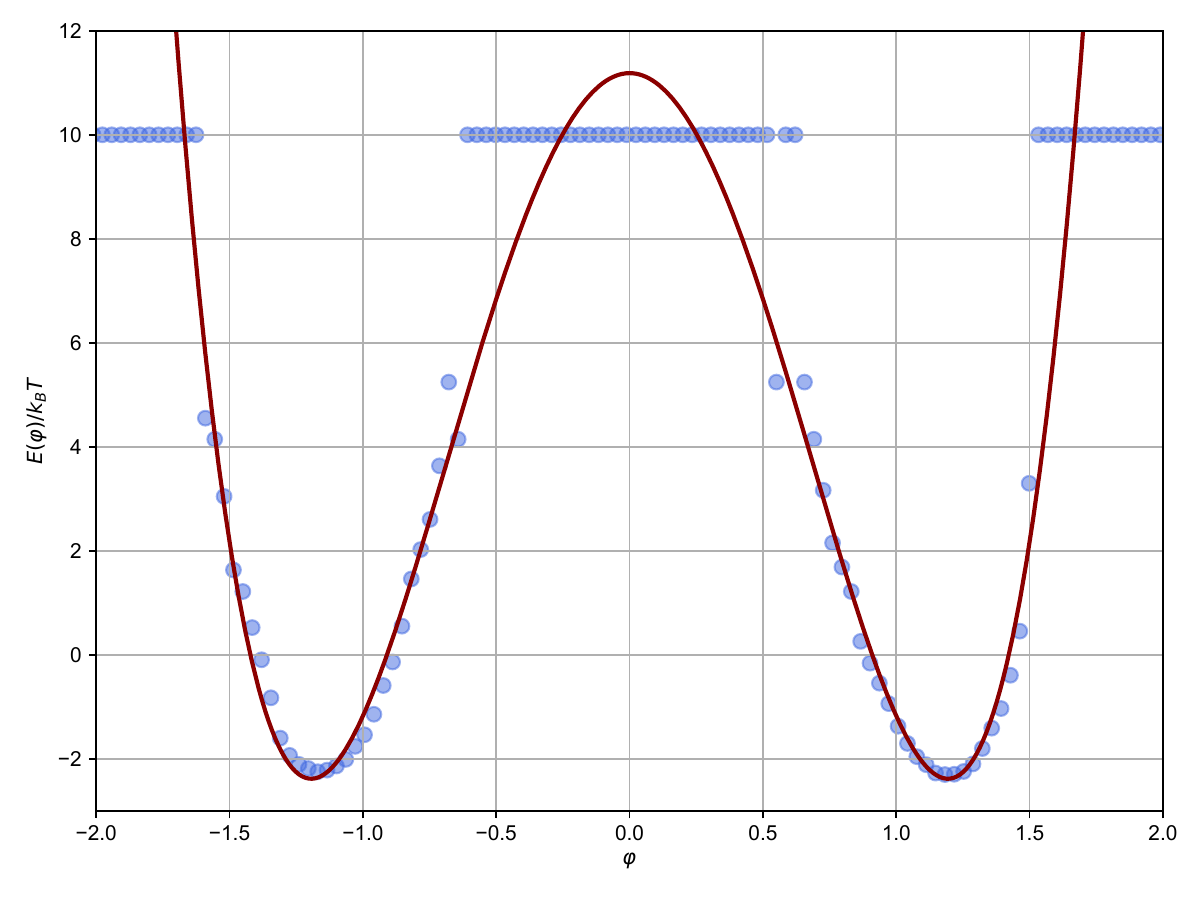}
\caption{A fit of the quartic potential to our data (fit: red line, data: blue points). Data points with $E = 10~k_B T$ correspond to angular positions where the pointer arm was never observed in the experiment and were excluded from the fit.}
\label{SI6}
\end{figure}

\subsection*{Kramers escape rate}
The escape rate from one of the minima across the barrier is given by
\begin{equation}
k_\mathrm{escape} = \frac{\sqrt{E''(\varphi_\pm) \mid E''(0) \mid}}{2\pi\gamma_r} \cdot e^{-\Delta E/k_B T} 
\label{Kramers}
\end{equation}

The second derivative of the potential energy is 
\[
  \frac{d^2 E}{d\varphi^2} = 12 A \varphi^2 - 2 B,
\]

hence
\[
E''(\varphi_\pm) = 12 A \varphi_\pm^2 - 2 B = 12 A \frac{B}{2 A} - 2B = 4 B
\]
and
\[
E''(0) = -2B
\]

Therefore, the prefactor becomes
\[
\frac{\sqrt{E''(\varphi_\pm) \mid E''(0) \mid}}{2\pi\gamma_r} = \frac{\sqrt{8B^2}}{2\pi\gamma_r} = \frac{\sqrt{2}B}{\pi\gamma_r},
\]

and the Kramers rate is: 
\[
k_\mathrm{escape} = \frac{\sqrt{2}B}{\pi\gamma_r} \cdot e^{-B^2/4 A k_B T} 
\]

\subsection*{External potential}

\begin{figure}[htbp!]
\centering
\includegraphics[width=0.8\textwidth]{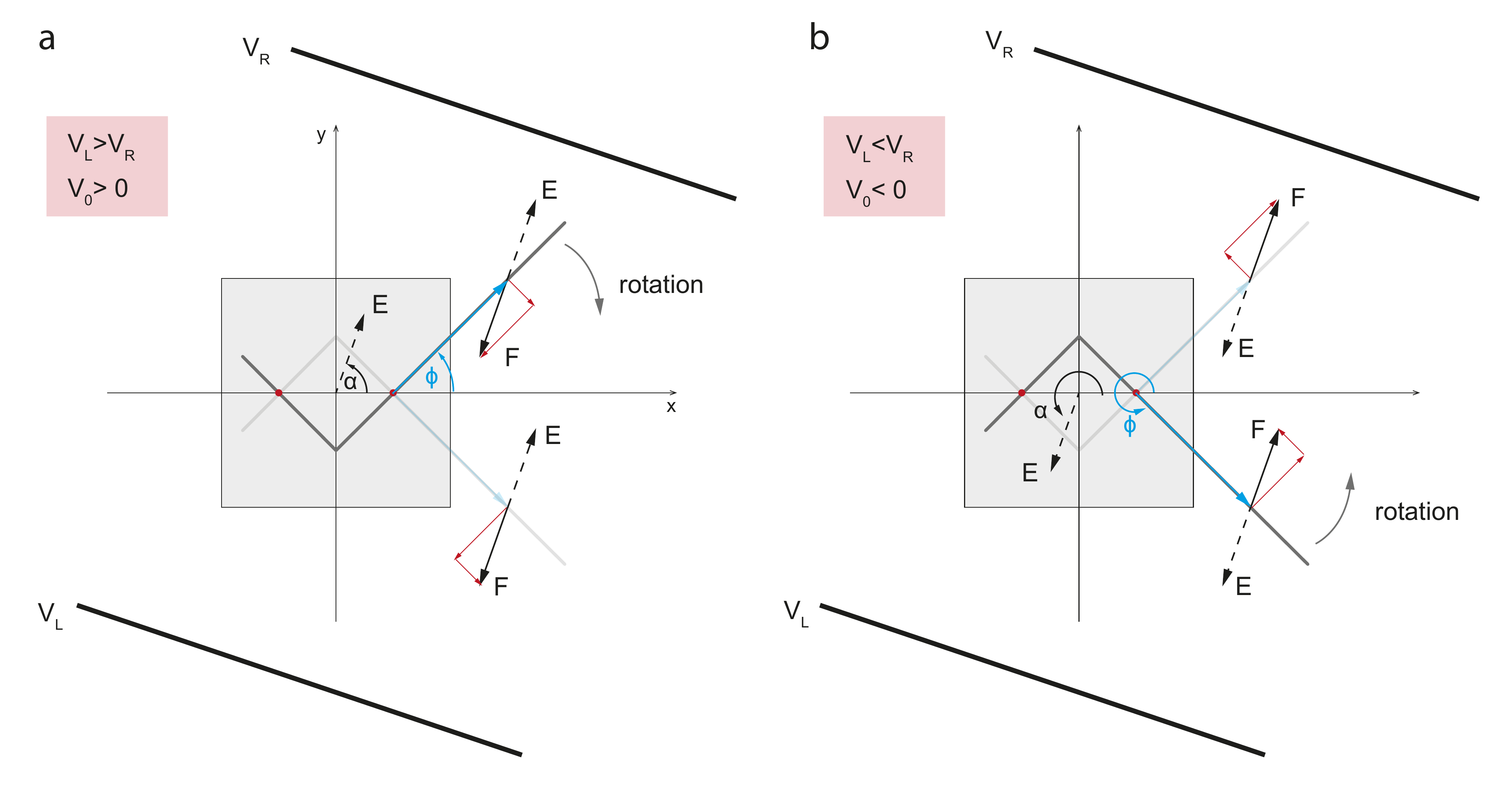}
\caption{Geometry of the experimental setup and angle definitions: $\varphi$ denotes the angle of the lever arm with respect to the $x$-axis, while $\alpha$ is the angle of the electric field direction. a) For positive voltages $V_0 > 0$, the negatively charged DNA lever arm is pulled towards the electrode with the more positive potential, and b), vice versa, for negative voltages $V_0 < 0$.}
\label{SI7}
\end{figure}

In the presence of an electric field, the components of the bistable switch experience a torque. We here focus on the 
long lever arms that are assumed to dominate the overall behavior. 
An external potential $V_0=V_L - V_R$ will create an electric field $\vec{E}$ in the sample chamber that exerts
a force on the charges of the lever arm. We assign an effective charge $-Q_\mathrm{eff}=- N_\mathrm{eff} \times e$ to the arm, which accounts for screening 
effects. Due to the negative charge of the DNA lever, the force will act in the direction of the positive electrode, i.e., in the opposite direction
than the electric field. 
We measure the orientations of the lever arm and the electric field with respect to the horizontal ($x$) axis of the platform with the angles $\varphi$, $\alpha$ (see Figure \ref{SI7}). The torque $\vec{\tau}$ exerted on the arm is given by 

\begin{align*}
\vec{\tau} &= \vec{r} \times \vec{F} 
= -Q_\mathrm{eff} \vec{r} \times \vec{E} \\
&= -Q_\mathrm{eff} E L \begin{pmatrix} \cos\varphi \\ \sin\varphi \\ 0 \end{pmatrix} 
   \times \begin{pmatrix} \cos\alpha \\ \sin\alpha \\ 0 \end{pmatrix} \\
&= -Q_\mathrm{eff} E L \begin{pmatrix} 0 \\ 0 \\ \sin\alpha \cos\varphi - \sin\varphi \cos\alpha \end{pmatrix} \\
&= -Q_\mathrm{eff} E L \begin{pmatrix} 0 \\ 0 \\ \sin(\alpha - \varphi) \end{pmatrix}
\end{align*}

Hence, 
\begin{eqnarray*}
\tau = |\vec{\tau}| =  -Q_\mathrm{eff} E L \sin(\alpha - \varphi) = -\frac{Q_\mathrm{eff} V_0 L}{d} \sin(\alpha - \varphi) = -\xi V_0 \sin(\alpha - \varphi),
\end{eqnarray*} 

Here, we simply set $E =V_0/d=(V_L-V_R)/d$, where $d$ is an effective electrode distance. The parameters $Q_\mathrm{eff}$ and $d$ are not really
known, and for simplicity all the prefactors have therefore been merged into the parameter $\xi$, which converts voltage
into an energy scale. Experimentally, we found that $\xi$ is typically on the order of $0.1~k_BT/\rm V$.

The moment $\tau$ can also be expressed in terms of a potential energy $E_\mathrm{el}(\varphi)$ via 
\[
 \tau = -\frac{dE_\mathrm{el}(\varphi)}{d\varphi}
\]

Thus
\[
E_\mathrm{el}(\varphi)=\xi V_0 \cos(\alpha - \varphi)
\]

This potential energy accounts for the externally applied electric field, which has to be added to the intrinsic (mechanical) energy landscape. 

\subsection*{Combined potential}
\begin{figure}[htbp!]
\centering
\includegraphics[width=0.8\textwidth]{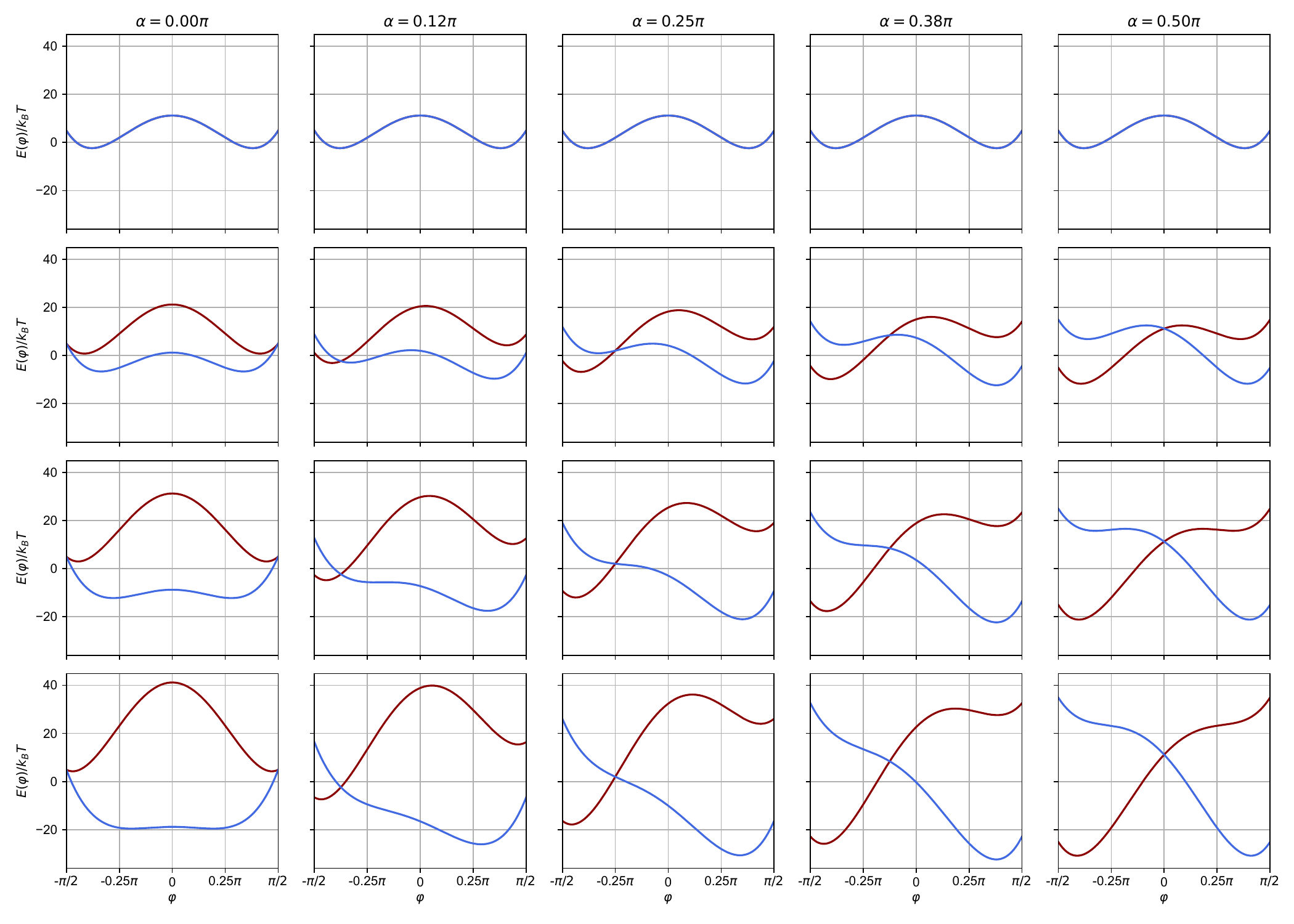}
\caption{Energy landscapes $E_\mathrm{tot}(\varphi, V_0)$ for positive ($V_0 > 0$, red) and negative($V_0 < 0$, blue) bias for different angles $\alpha$. For some angles, switching in one direction is favored, but disfavored in the other. For $\alpha = \pi/2$ forward and backward switching are favored symmetrically.}
\label{SI8}
\end{figure}

Hence, the total potential is:
\[
E_\mathrm{tot}(\varphi)=A \varphi^4 - B \varphi^2 + C +  \xi V_0 \cos(\alpha - \varphi)
\]

For increasing $V_0$, depending on the angle, the potential is tilted, resulting in altered minima, maxima and energy barrier.
For a given set of parameters $A,B,C$ (determined from the zero-field experiments) and $\alpha,V_0$ we can calculate the minima and maxima as well as the local curvatures of the total potential. From these we obtain the forward and backward energy barriers $\Delta E_{f,b}$ as well as the transition rates $k_+(\alpha,V_0)$ and $k_-(\alpha,V_0)$ between the two potential wells. There are situations, in which the field promotes transitions in one direction, but cannot switch the structure back, when reversing the field.  
We determined the "optimum" switching angle as the angle for which the \emph{sum} of the forward energy barrier $\Delta E_f (\alpha, V_0)$ and 
the backward energy barrier for the reverse field direction $\Delta E_b (\alpha, -V_0)$  is minimized, which gives $\alpha = \pm \pi/2$.

\begin{figure}[htbp!]
\centering
\includegraphics[width=0.5\textwidth]{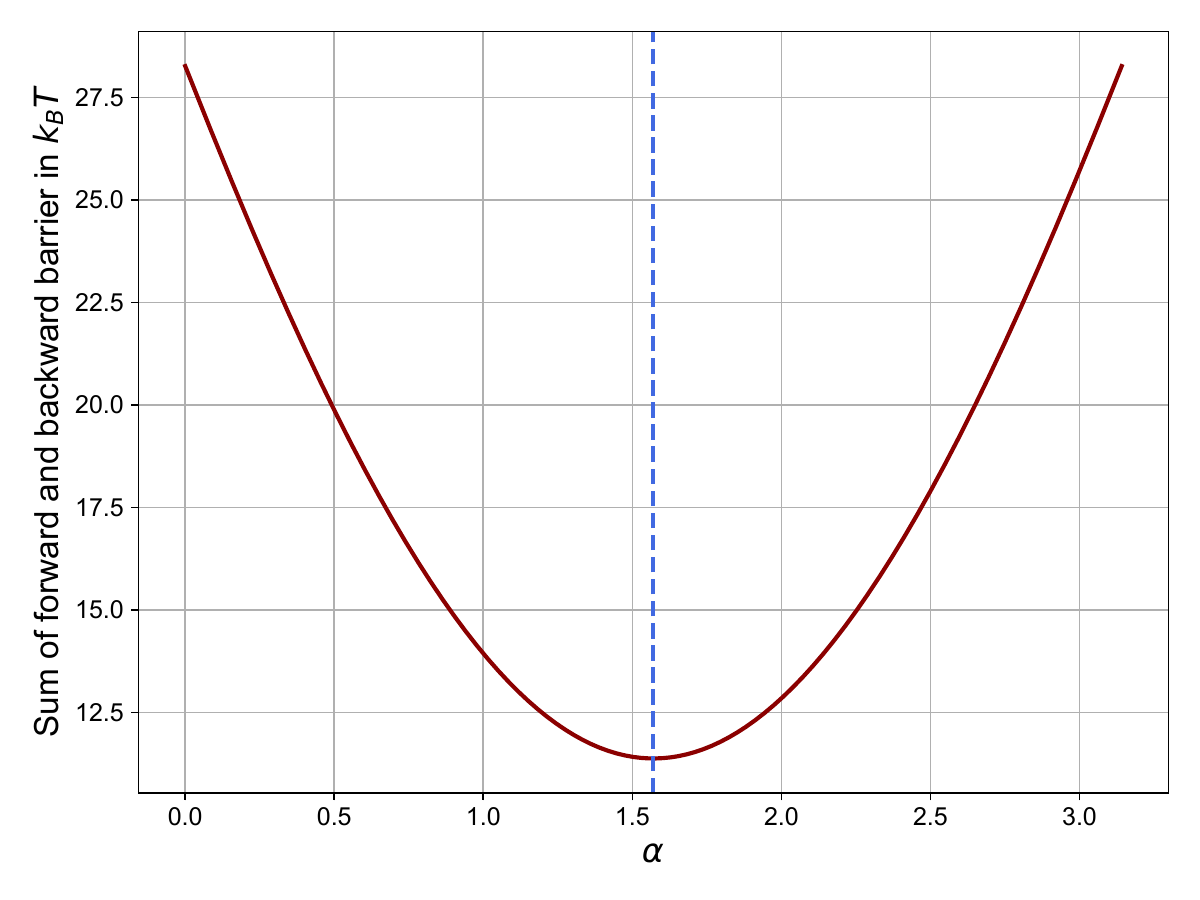}
\caption{Sum of the energy barrier in forward direction and the corresponding barrier in backward direction after reversal of the voltage. Optimal switching conditions correspond to a minimum of the sum, which occurs for an angle of $\alpha=\pi/2$.}
\label{SI9}
\end{figure}

\subsection*{Escape rate as a function of voltage}
For the following we focus on the optimum angle $\alpha = \pi/2$. For large
barriers ($\Delta E \gtrsim 2 k_BT$), we can estimate the escape time from one of the minima using Kramers' formula (Eq.~\ref{Kramers}). However,
this formula becomes innacurate for low barriers and shallow minima. 
To obtain a more accurate estimate for the escape rate for smaller barriers, we determine the mean first passage time (MFPT) for
the transition from the minimum to the barrier:

\[
\text{MFPT} = \frac{\gamma_r}{k_B T} \int_{\varphi_0}^{\varphi_b} dx \, e^{V(x)/k_B T} \int_{-\infty}^{x} dy \, e^{-V(y)/k_B T}
\]

\begin{figure}[htb]
\centering
\includegraphics[width=\textwidth]{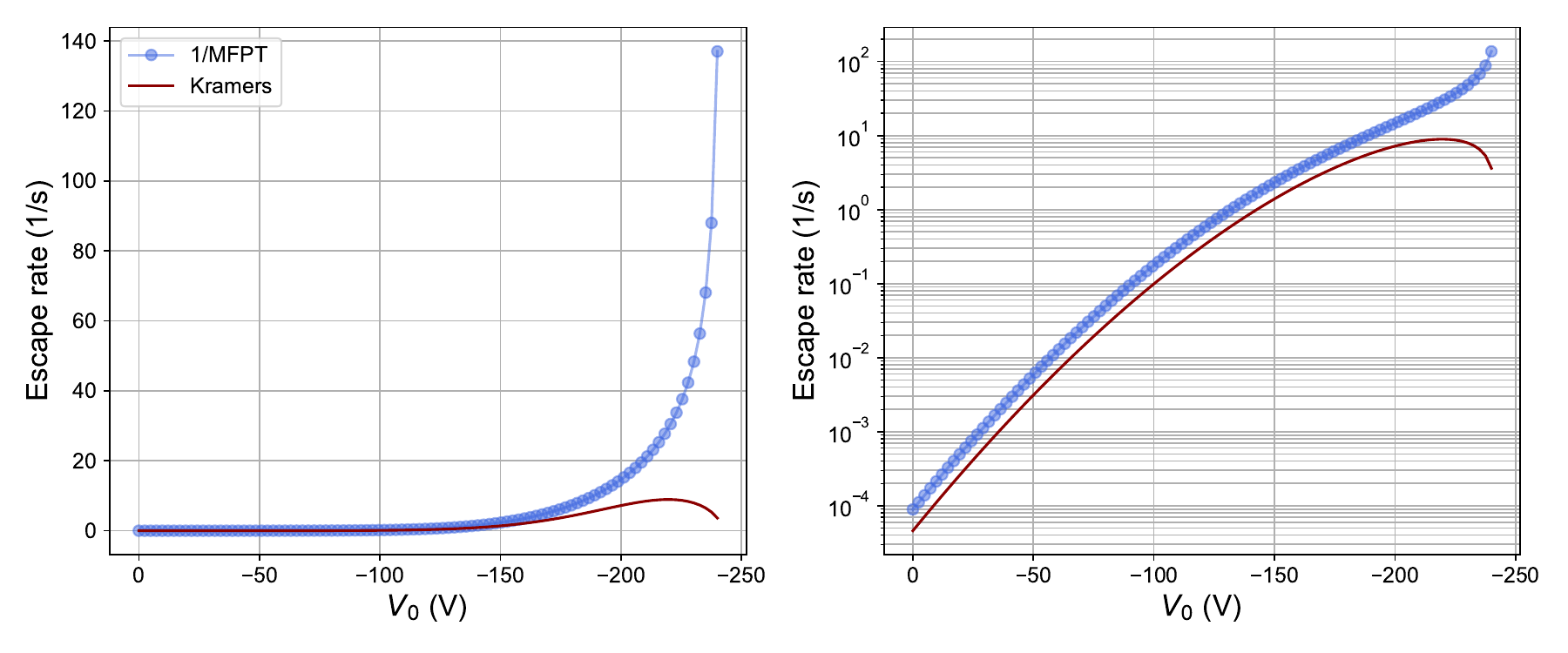}
\caption{Escape rates from the potential minimum are shown for different applied bias voltages $V_0$. The plots are presented as a linear plot (left) and a log-linear plot (right). The blue curve represents the escape rate calculated from the mean first passage time (MFPT) in the biased double-well potential, evaluated from the minimum to the position of the energy barrier. The red curve shows the corresponding estimate based on Kramers' approximation. Kramers' approximation breaks down for low energy barriers and underestimates the escape rate compared to the MFPT calculation. The energy barrier vanishes at approximately $V_0 \approx 240\,\mathrm{V}$, beyond which the MFPT approach also becomes inaccurate. All plots are computed for the optimal switching angle $\alpha = \pi/2$.}
\label{SI10}
\end{figure}

A comparison of the Kramers prediction and the rate derived from the MFPT are shown in Figure \ref{SI10}.

We can now estimate the switching probability by considering the probability that the switch escapes from a local energy minimum after 
reversing the electric field for a pulse length of $t_p$, which is simply given by 

\[
p_\mathrm{switch}=1-\exp{\left(-k t_p\right)}
\]

\begin{figure}[htbp!]
\centering
\includegraphics[width=0.6\textwidth]{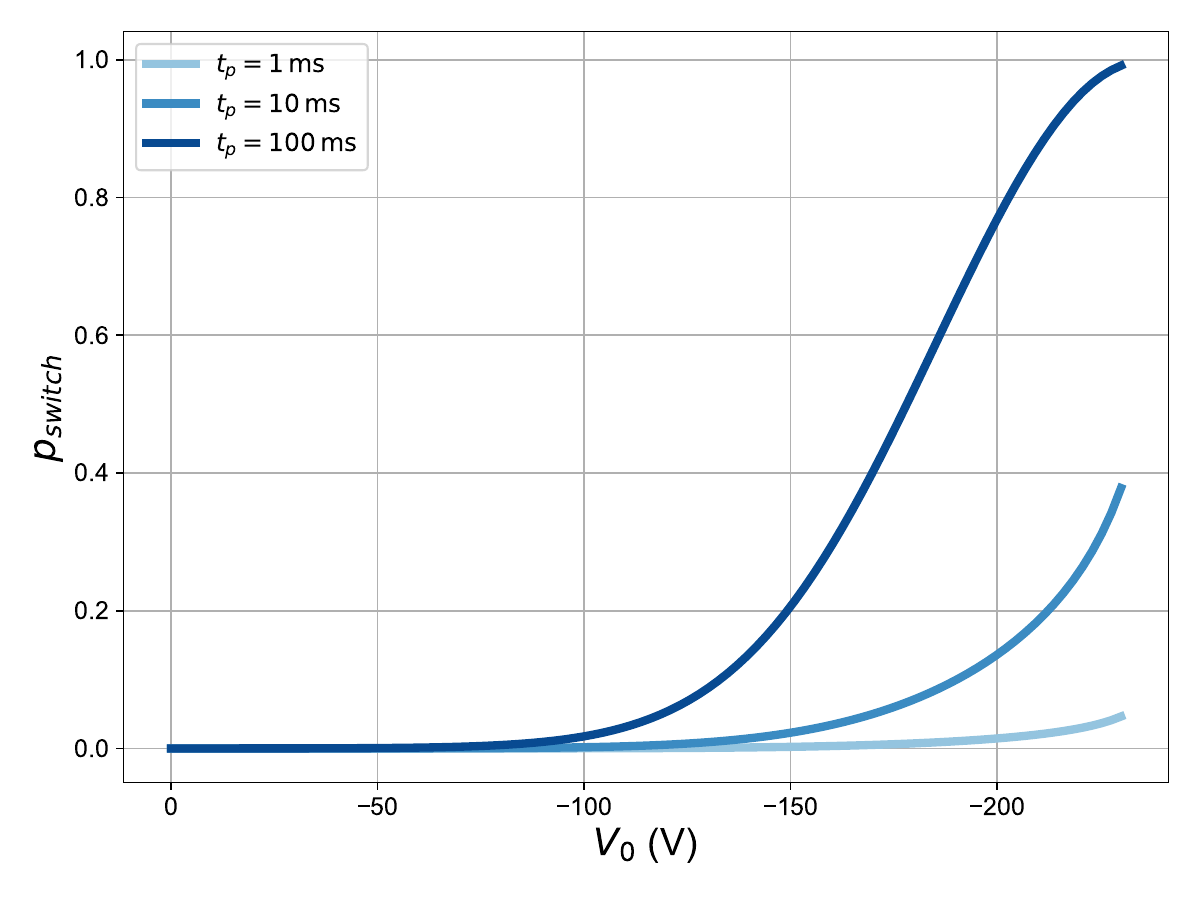}
\caption{Escape probability from a potential minimum after switching the electric field for a given pulse length $t_p$, derived from the MFPT rate.}
\label{SI11}
\end{figure}

A plot of $p_\mathrm{switch}$ for the experimentally used pulse lengths $t_p = 1~\rm ms, 10~ms, 100~ms$ is shown in Figure \ref{SI11}. 

Our MFPT approach becomes inaccurate when reaching voltages for which the barrier vanishes altogether (which occurs around $V_0 = 240 V$ in the model for $\alpha = \pi/2$), and the pointer arm simply drifts in the tilted potential. 
The drift time from $\varphi_a$ to $\varphi_b$ is given by: 

\[
T_\mathrm{drift} = - \gamma_r \int_{\varphi_a}^{\varphi_b} \left(\frac{dV}{d\varphi}\right)^{-1} d\varphi
\]

For large voltages, we approximate  $\left(\frac{dV}{d\varphi}\right)^{-1} \approx -1/(\xi V_0)$, and thus 

\[
T_\mathrm{drift} \approx \frac{\gamma_r \Delta \varphi}{\xi V_0}
\]

For our estimated parameters $\gamma_r = 0.24~k_B T \rm s$ and $\xi = 0.1~k_B T/\rm V$ this becomes $\approx 1~\rm ms$ for $V_0 = 240~\rm V$.

\subsection*{Mechanical considerations for snap-through mechanism}
Our OxDNA simulations (section 2) indicate that a combination of effects - helix fraying, bending, and out-of-plane movement - is involved in the snap-through transition. Interestingly, as discussed below, for none of the isolated effects alone the estimated forces/energies are within the correct order of magnitude.

\subsubsection*{Beam-bending}
The persistence length of a DNA origami N-helix bundle is given by 
\begin{equation}
L_p^N =N L_p \left(1 + \frac{2}{\sin^2{\pi/N}} \right)
\end{equation}

The energy required to bend the semirigid bundle with length $L$ and persistence length $L_p^N$ into curved segment with radius of curvature $R$ is given by
\begin{equation}
E_\mathrm{bend}= \frac{k_BT L_p}{2}\int_0^{L} \frac{1}{R^2} ds = \frac{L L_p^N}{2 R^2} \times k_B T 
\end{equation}

For $N=6$, with $L_p = 50~\rm nm$, we obtain $L_p^6 = 2.7~\rm \mu m$. 

The length of the rotor arms from their respective 
rotor hinges to the elbow hinge is $L \approx 15.9 \rm nm$, whereas the distance between the rotor hinges is $d_\mathrm{hinge}$. 
We can estimate the energy cost for bending the beams such that they would fit through the limited length $d_\mathrm{hinge}$.
When a beam of length $L$ is bent into  an arc with chord length $c = d_\mathrm{hinge}/2$, its radius of curvature $R$ can be determined in the following way:

From
\[
L = R \cdot \theta
\]

and 

\[
c = 2R\cdot \sin\left( \frac{\theta}{2} \right)
\]

we get
\[
d_\mathrm{hinge} = 2 c = 4R \cdot \sin\left( \frac{L}{2R} \right).
\]

This expression can be numerically solved for $R$, from which one can determine the bending energy $E_\mathrm{bend}$. For our geometry this energy is $\approx 1,100 \rm k_B T$, two orders of magnitude too high! 

\subsubsection*{Buckling}
Instead of bending, the beam might buckle under the exerted forces. The expression for the critical Euler buckling force $F_{\text{crit}}$ of a semiflexible polymer
is given by

\[
F_{\text{crit}} = \frac{\pi^2 k_B T L_p}{L^2}
\]

For an N-helix bundle, replacing $L_p$ with $L_p^N$, this force is on the order of $500~\rm pN$, which again is two orders of magnitude too high.

\subsubsection*{Force-induced melting}
A potential alternative mechanism facilitating bistable switching could be the partial melting/fraying of DNA double helical sections of the DNA origami structure, which would lead to a softening of the structure and thus a reduction of the energy barrier.
Such a melting mechanism has also been discussed in the context of the anomalously high cyclization efficiencies observed for short dsDNA~\cite{Yan:2004vt}. 

Depending on the mode of force application, the force required to induce melting of double-stranded DNA varies significantly, however. When unzipping the strands in a zipper-like manner, typical melting forces range from \SIrange{10}{15}{\pico\newton}, depending on the base-pair sequence. In contrast, axial stretching of the helix leads to an overstretching/melting transition around \SI{65}{\pico\newton}. Applying shear forces—by pulling opposite strands in opposite directions—requires much higher forces of order \SI{150}{\pico\newton}. 
\clearpage

\bibliographystyle{naturemag}
\bibliography{scibib}